\documentclass[twocolumn,tighten]{aastex63}
\usepackage{graphicx}
\usepackage{amsmath}
\usepackage{natbib}
\usepackage{amssymb}

\usepackage{sidecap}
\usepackage{lineno}
\def\KB{k_{\rm B}}
\def\MP{m_{\rm p}}
\def\Msun{{\mathrm{M}_\odot}}
\def\Rsun{{\mathrm{R}_\odot}}

\shorttitle{Light Curve Model for Luminous Red Novae}
\shortauthors{Matsumoto \& Metzger}

\begin{document}

\title{Light Curve Model for Luminous Red Novae and Inferences about the Ejecta of Stellar Mergers}

\author{Tatsuya Matsumoto}
\affil{Department of Physics and Columbia Astrophysics Laboratory, Columbia University, Pupin Hall, New York, NY 10027, USA}

\author[0000-0002-4670-7509]{Brian D. Metzger}
\affil{Department of Physics and Columbia Astrophysics Laboratory, Columbia University, Pupin Hall, New York, NY 10027, USA}
\affil{Center for Computational Astrophysics, Flatiron Institute, 162 5th Ave, New York, NY 10010, USA}

\begin{abstract}
The process of unstable mass transfer in a stellar binary can result in either a complete merger of the stars or successful removal of the donor envelope leaving a surviving more compact binary.  ``Luminous red nova'' (LRN) are the class of optical transients believed to accompany such merger/common envelope events.
Past works typically model LRNe using analytic formulae for supernova light curves which make assumptions (e.g., radiation dominated ejecta, neglect of hydrogen recombination energy) not justified in stellar mergers due to the lower velocities and specific thermal energy of the ejecta.  We present a one-dimensional model of LRN light curves, which accounts for these effects.
Consistent with observations, we find that LRNe typically possess two light curve peaks, an early phase powered by initial thermal energy of the hot, fastest ejecta layers and a later peak powered by hydrogen recombination from the bulk of the ejecta.  We apply our model to a sample of LRNe to infer their ejecta properties (mass, velocity, and launching radius) and compare them to the progenitor donor star properties from pre-transient imaging.  We define a maximum luminosity achievable for a given donor star in the limit that the entire envelope is ejected, finding that several LRNe violate this limit.  Shock interaction between the ejecta and pre-dynamical mass-loss,
may provide an additional luminosity source to alleviate this tension.
Our model can also be applied to the merger of planets with stars or stars with compact objects.   
\end{abstract}


\section{Introduction}
The direct interaction between the stars in a binary system is a common occurrence (e.g., \citealt{Sana+12}), with implications for a number of topics in stellar evolution (e.g., \citealt{Podsiadlowski+04,deMink+14,DeMarco&Izzard17}) and energetic transients (e.g., \citealt{Soker&Tylenda06,Chevalier12,Smith14}).  Such interaction often begins after the (typically more evolved) donor star overflows its Roche lobe and begins transferring mass onto the more compact accretor star.  Depending on the binary mass ratio and eccentricity and the evolutionary state of the donor, this mass-transfer can become unstable, resulting in a runaway growth of the mass transfer rate (e.g., \citealt{Soberman+97,Pavlovskii+17,Ge+20,Klencki+21}).  

Following an initial period of mass-loss, potentially lasting many orbits and likely concentrated in the binary plane (e.g., from the $L_2$ Lagrange point; \citealt{Pejcha+16a,Pejcha+16b,Pejcha+17,MacLeod&Loeb20}) it is commonly believed that the accretor will ultimately be dragged into the envelope of the donor star \citep{Paczynski76,Taam+78,Iben&Livio93,Ivanova+13a}.  The final outcome of such a ``common envelope'' phase can be either a complete merger of both stars (e.g., \citealt{Kruckow+16}), or the successful ejection of the donor envelope and the survival of a tighter binary comprised of the accretor and the core of the original donor (e.g., \citealt{Ivanova&Nandez16,Sand+20,Law-Smith+20}). 

Stellar mergers and/or common envelope events have been invoked to give rise to number of atypical stellar and binary populations, including cataclysmic variables \citep{Patterson84}, magnetic A-stars \citep{Schneider+19}, magnetic white dwarfs \citep{Tout+08,Nordhaus+11}, sdB-stars (e.g., \citealt{Kramer+20}), carbon-deficient giants (e.g., \citealt{Bond19}), and blue stragglers (e.g., \citealt{Davies+04,Ferreira+19,Wang+20}), among other examples.  The conditions required to remove the envelope, and the final separations of the surviving binary, are also important for predicting the population of merging binary compact objects by LIGO/Virgo (e.g., \citealt{Tauris+17,Belczynski+18,Vigna-Gomez+18}).  However, despite increasingly sophisticated multi-dimensional simulations of the common envelope phase over the past decade (e.g., \citealt{Ricker&Taam12,Ohlmann+16,Ivanova&Nandez16,MacLeod+18,Prust&Chang19,Chamandy+19,Law-Smith+20}), the theoretical community has yet to converge on a consistent physical picture which includes all of the relevant physical processes \citep{Clayton+17,Soker+18,Wilson&Nordhaus19,Reichardt+20}.  Observations remain extremely important to unraveling this mystery.

\begin{figure*}
\begin{center}
\includegraphics[width=160mm, angle=0,bb=0 0 339 341]{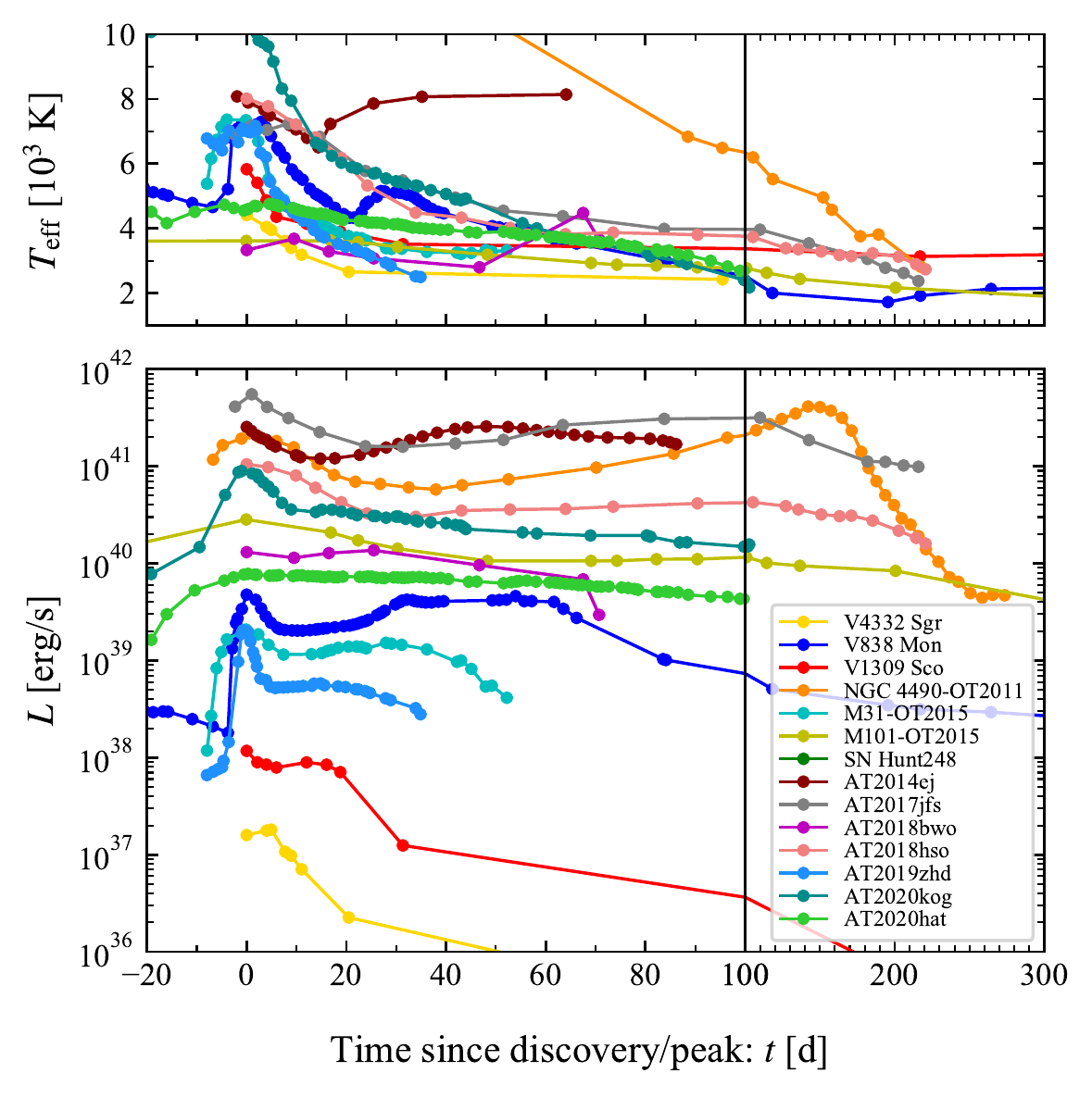}
\caption{Bolometric light curves (bottom panel) and effective temperatures (top panel) as a function of time since peak or discovery for a sample of LRNe. See Table \ref{table data} for references to the data used to construct the figure. }
\label{fig:obs}
\end{center}
\end{figure*}

The dynamical merger event may generate a short-lived transient powered by the ejection of mass, primarily from the donor's envelope \citep{Soker&Tylenda06}.  ``Luminous red novae'' (LRN) are the class of optical transients with typical durations of weeks to months and peak luminosities $\sim 10^{38}-10^{41}$ erg s$^{-1}$ between those of classical novae and supernovae (SNe) and characteristically red colors (e.g., \citealt{Martini+99,Munari+02,Bond+03,Kulkarni+07,Tylenda+11,Kurtenkov+15,Smith+16,Blagorodnova+17,Cai+19,Stritzinger+20}; see \citealt{Pastorello+19a} for a recent review and Fig.~\ref{fig:obs} for a sample of LRNe).    The watershed event which confirmed the association between LRNe and merging binary stars was the Galactic source V1309 Sco \citep{Mason+10}, detected by OGLE as a contact binary for almost a decade prior to the outburst \citep{Tylenda+11}.  The observed evolution of the binary light curve and orbital period leading up to the merger, provided evidence for high rates of mass-loss, starting hundreds of orbits prior to the final dynamical coalescence and concomitant abrupt rise to optical maximum \citep{Pejcha14,Pejcha+17,MacLeod&Loeb20}.  Similar ``precursor" emission, potentially powered by internal shocks in outflowing streams from the $L_2$ point (e.g., \citealt{Pejcha+16a,Pejcha+16b}), is observed in the light curves of other LRNe, suggesting that pre-dynamical mass loss is common if not ubiquitous in stellar mergers (\citealt{Metzger&Pejcha17,Blagorodnova+21}).  

\begin{figure*}
\begin{center}
\includegraphics[width=160mm,angle=0,bb=0 0 440 214]{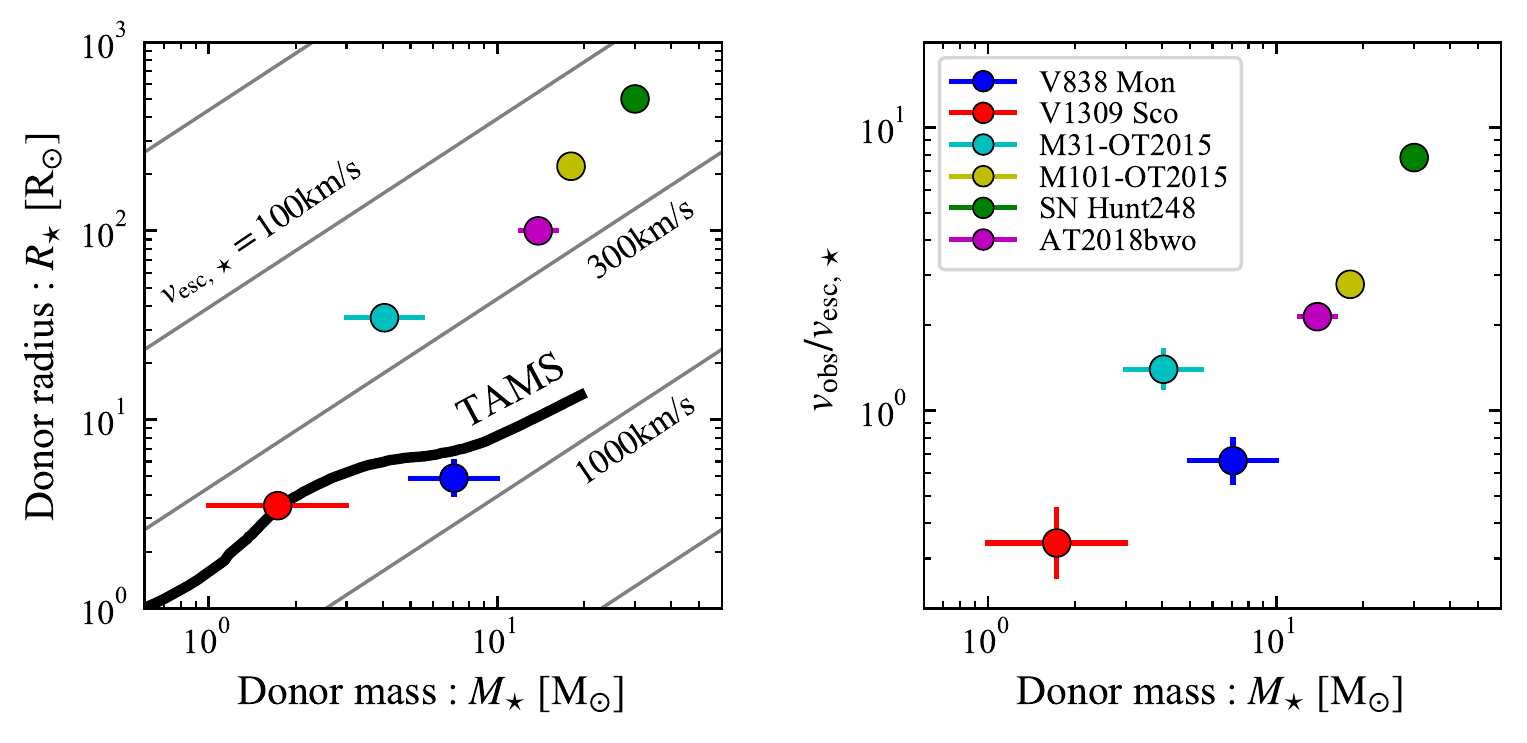}
\caption{(Left Panel) Progenitor star $R_{\star}$ versus progenitor star mass $M_{\star}$ as inferred from pre-transient observations for a sample of LRNe (Table \ref{table data}).  For comparison we show the terminal age main sequence (TAMS; black line) and contours of surface escape velocity $v_{\rm esc,\star} = (2GM_{\star}/R_{\star})^{1/2}$ (gray lines). (Right Panel) Ratio of the ejecta velocity $v_{\rm obs}$ inferred from spectroscopy of the transient emission to progenitor donor surface escape speed $v_{\rm esc,\star}$.  As expected, $v_{\rm obs}/v_{\rm esc,\star}$ is larger for more evolved progenitors ($R_{\star} > R_{\rm TAMS})$ with well-developed core-envelope structure, for which the ejecta launching radius $R_0$ can be $\ll R_{\star}$.} 
\label{fig:progenitor}
\end{center}
\end{figure*}

The light curves of LRNe encode information on the merger ejecta, particularly its mass, velocity, and launching radius (e.g., \citealt{Ivanova+13b,MacLeod+17,Blagorodnova+21}).  Accurate determinations of these properties can, in combination with information on the progenitor binary from pre-outburst imaging, be used to constrain the physical processes at work leading up to and during the merger (e.g., \citealt{MacLeod+17,Blagorodnova+21}).  \citet{Ivanova+13b} pioneered this approach by applying an analytic model for the luminosity and duration of Type II SNe (SNe II) light curves from \citet{Popov93} to infer the ejecta properties for a sample of LRNe.  Using an updated version of the \citet{Popov93} formulae calibrated using SNe II radiative transfer calculations \citep{Sukhbold+16}, \citet{Blagorodnova+21} constrained the ejecta mass and launching radius for the AT2018bwo.  Combined with an application of binary stellar evolution models to the observed yellow supergiant progenitor, the ejecta properties enabled them to paint a consistent picture of the event (see also \citealt{MacLeod+17} for a similar analysis of M31-OT2015).  Most LRN progenitor stars are moderately evolved, with radii between $\sim 1-10$ times that of main sequence stars of their mass (Fig.~\ref{fig:progenitor}).

Despite the advances of these past works, several physical inconsistencies can arise when applying SNe II models to LRN.  SNe ejecta are dominated by radiation pressure, an assumption implicit in SN II light curve models \citep{Popov93,Faran+19}.  By contrast, given the comparatively lower velocities, higher densities, and lower thermal energies of LRN ejecta, gas pressure can dominate over radiation pressure and sources of opacity other than electron scattering may be important.  Furthermore, while hydrogen recombination could be an important source of luminosity for LRN (e.g., \citealt{Ivanova+13b}), recombination energy is irrelevant in SNe and is not included in models like \citet{Popov93}.  SN light curves are powered by the initial thermal energy of the recently-shocked ejecta from the explosion or the radioactive decay of $^{56}$Ni.  Although the early-time peaks seen in many LRN light curves can be powered by the initial thermal energy of hot ejecta \citep{MacLeod+17,Metzger&Pejcha17}, the longer-lived plateau emission (sometimes manifesting as a second luminosity peak) is powered by hydrogen recombination energy or other forms of radially-distributed ejecta heating (e.g., shock interaction between the fast merger ejecta and circumbinary material from the pre-dynamical phase; \citealt{Metzger&Pejcha17}, or from an accretion-powered jet; e.g., \citealt{Soker20,Soker&Kaplan21}).

In this paper we present a new model for LRN light curves, which self-consistently accounts for the effects of arbitrary gas and radiation pressure, thermal energy released by recombination, and a realistic distribution of ejecta velocities.  After presenting the light curve model in Section \ref{sec:method} we describe our results in Section \ref{sec:results}.  After walking the reader through a few example light curves (Sec.~\ref{sec:example}) and elucidating the physics giving rise to multiple light curve peaks, we perform a parameter study covering the full range of stellar merger properties (Sec.~\ref{sec:parameterstudy}), apply our model to extract the ejecta properties from a sample of individual LRNe (Sec.~\ref{sec:IndividualLRN}), and compare the predictions of our light model to findings obtained using SN II models (Sec.~\ref{sec:Popovcompare}).  Section \ref{sec:discussion} discusses implications of our findings, including the need for alternative energy sources in LRN from circumbinary interaction (Sec.~\ref{sec:shocks}); the populations of binaries giving rise to LRNe (Sec.~\ref{sec:LRNpopulation}); and other applications of our models to planet-star or star-compact object mergers (Sec.~\ref{sec:othermergers}).  We briefly summarize our conclusions in Section \ref{sec:conclusions}.

\section{Light Curve Model}
\label{sec:method}

Our light curve model follows that developed by \citet{Metzger+21} in the different context of optical transients from the tidal disruption of a star by a white dwarf, but with added physical effects such as radiation pressure and Helium ionization \citep{Kasen&RamirezRuiz10}.  The general approach is to divide the radially-stratified ejecta into successive thin shells, calculate the emission from each such mass/velocity shell separately by means of a time-dependent one-zone model, and then obtain the total light curve by summing over the contributions from each shell.

We begin by describing the one-zone model, as will ultimately be applied to each shell of velocity $v$ and external mass $M_{\rm ej}(>v)$. Immediately after the stellar merger or common envelope event, the evolution of the unbound ejecta will be complex, potentially involving self-interaction between different ejecta components (e.g., \citealt{Metzger&Pejcha17,MacLeod+18,Reichardt+19}).  However, we may reasonably assume that on a few dynamical times the outflow geometry will become quasi-spherical and the thermal and kinetic energies roughly equal. Hence we set the initial internal energy of the shell as $E_0=dMv^2/2$ (here $dM$ is the shell's mass) and calculate the subsequent evolution starting from time $t_0=R_0/v$, where $R_0$ is the initial (``ejection'') radius of the ejecta (typically equal to, or smaller than, the radius of the donor star).  After the initial thermal energy has been converted into kinetic energy on a timescale $\sim t_0$, the shell will thereafter expand freely with radius $R=vt$.  

The internal energy of the shell can be written
\begin{align}
E&=\frac{3}{2}(1+\bar{x})N\KB T+aT^4V+\sum_{i}NA_ix_i\varepsilon_{i}\ ,
	\label{eq:internal_energy}
\end{align}
where $\bar{x}=\sum_i A_ix_i$, $A_i$, $x_i$, and $\varepsilon_i$ are the mean ionization degree, abundance fraction, degree of ionization, and ionization energy of species $i$, respectively.  We calculate $\bar{x}$ and $x_{i}$ by solving Saha equation taking into account singly-ionized hydrogen and helium, which we have checked capture recombination effects on the light curve to high accuracy.  The other quantities are the total number of nuclei $N$, the Boltzmann constant $\KB$, temperature $T$, radiation constant $a$, and shell volume $V=4\pi R^2dR$ with the shell width $dR$.

The internal energy evolves according to the first law of thermodynamics:
\begin{align}
\frac{dE}{dt}=-\frac{3(\gamma_3-1)E}{t}-L\ ,
    \label{eq:evolution}
\end{align}
where the effective adiabatic index is given by a function of density $\rho=dM/V$ and temperature $T$ \citep{KrishnaSwamy61,Kasen&RamirezRuiz10}
\begin{align}
\gamma_3(\rho,T)&\simeq1+\frac{(1+\bar{x})(1+4\alpha)+\sum_iA_i\frac{\bar{x}(1-\bar{x})}{(2-\bar{x})}\bigl(\frac{3}{2}+\frac{\varepsilon_{i}}{\KB T}\bigl)}{(1+\bar{x})\big(\frac{3}{2}+12\alpha\big)+\sum_iA_i\frac{\bar{x}(1-\bar{x})}{(2-\bar{x})}\bigl(\frac{3}{2}+\frac{\varepsilon_{i}}{\KB T}\bigl)^2}\ ,
    \label{eq:gamma3}
\end{align}
and
\begin{align}
\alpha&\equiv\frac{P_{\rm rad}}{P_{\rm gas}}=\frac{aT^3V}{3(1+\bar{x})N \KB}\ .
\end{align}
is the ratio of radiation pressure $P_{\rm rad} = aT^{4}/3$ to gas pressure $P_{\rm gas} = N(1+\bar{x})\KB T/V$.  The radiative luminosity of the shell is given by
\begin{align}
L = \frac{E_{\rm rad}}{t_{\rm d}+t_{\rm lc}}\ ,
    \label{eq luminosity}
\end{align}
where $E_{\rm rad}=aT^4V$ and the photon diffusion time through the column associated with the external mass is
\begin{align}
t_{\rm d}&=\frac{\kappa M_{\rm ej}(>v)}{4\pi cR}\ ,
    \label{eq:tdiff}
\end{align}
and the factor $t_{\rm lc}=R/c$ limits the energy escape time to the light crossing time, where $c$ is the speed of light. It would be more accurate to express Eq.~\eqref{eq:tdiff} as an integral over the diffusion time through the external layers $>v$ accounting for their respective opacity $\kappa$ and density distribution.  However, this approach would require following energy transport between the shells via radiative diffusion; such a detailed treatment is beyond the scope of this paper. Hence we adopt this simpler single-zone expression for $t_{\rm d}$ because we find the calculated light curve turn outs to be relatively insensitive to this assumption (a factor of 10 change in the pre-factor of Eq.~\eqref{eq:tdiff} changes the light curve duration by a factor $\lesssim 2$).  

\begin{figure}
\begin{center}
\includegraphics[width=85mm,angle=0,bb=0 0 280 210]{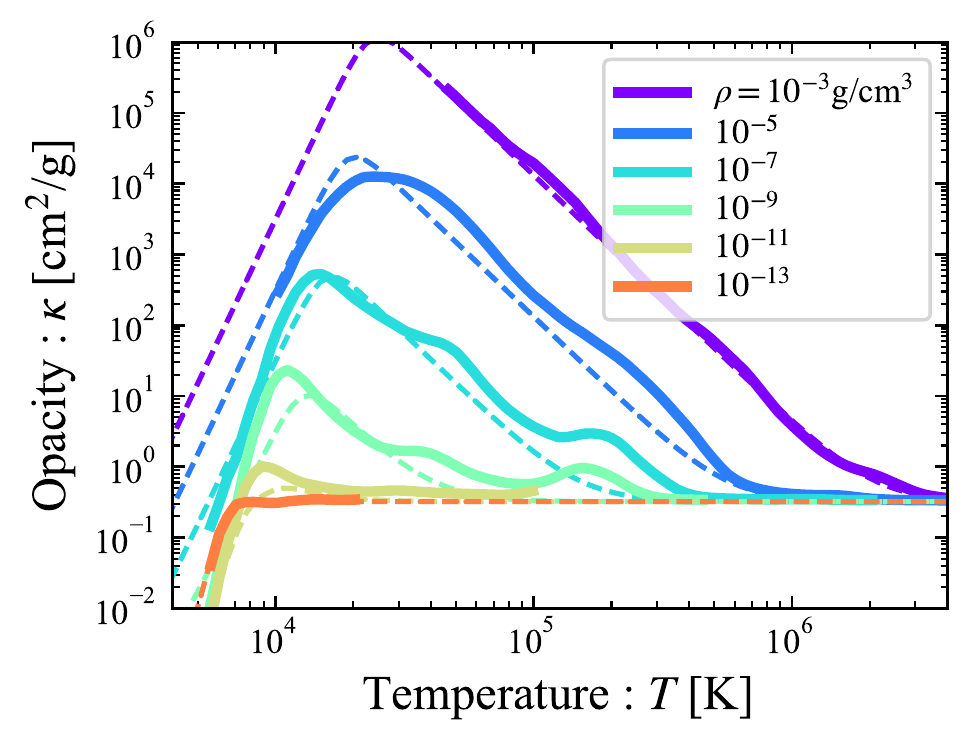}
\caption{Comparison of our assumed opacity law (dashed curve; Eq. \ref{eq opacity}) as a function of gas temperature and density with results from the OPAL opacity code (solid curves) for solar metallicity gas.}
\label{fig opacity}
\end{center}
\end{figure}

We approximate the Rosseland mean opacity using the analytic formula similar to that used by \citep{Metzger&Pejcha17}
\begin{align}
\kappa=\kappa_{\rm m}+\kappa_{\rm e}+\Big(\kappa_{\rm H^{-}}^{-1}+\kappa_{\rm K}^{-1}\Big)^{-1}\ ,
    \label{eq opacity}
\end{align}
which accounts for the contribution from molecular opacity $\kappa_{\rm m}\simeq0.1 \,Z\,\rm cm^2\,g^{-1}$, H$^{-}$ opacity $\kappa_{\rm H^{-}} \simeq 1.1\times10^{-25}\,Z^{0.5}\rho^{0.5}T^{7.7}\,\rm cm^2\,g^{-1}$, Kramers (bound-bound and bound-free) opacity $\kappa_{\rm K} \simeq 1.2\times10^{26}\,Z(1+X)\rho T^{-7/2}\,\rm cm^2\,g^{-1}$, and electron scattering $\kappa_{\rm es}\simeq0.4\,\bar{x}\mu^{-1}\,\rm cm^2\,g^{-1}$.
Here $Z$ is the metallicity, $X$ is the mass fraction of hydrogen, and $\mu \equiv dM/m_{\rm p}N$($\simeq1.25$ for the solar metallicity) is the mean molecular weight neglecting the number of electrons, where $m_{\rm p}$ is the proton mass.  We neglect dust opacity, which will become important once the dense ejecta cools below the temperature for solid condensation; our model cannot therefore be applied to the late phases of LRN emission when dust formation is seen to occur (e.g., \citealt{Wisniewski+08,Nicholls+13,Banerjee+15,Kasliwal+17}).  Figure~\ref{fig opacity} compares our assumed opacity law (Eq.~\ref{eq opacity}) to that obtained using the OPAL opacity code for $Z = Z_{\odot} \simeq 0.02$.\footnote{\texttt{https://opalopacity.llnl.gov/existing.html\#type1GN}} Note that our prescription of the electron scattering opacity in Eq.~\eqref{eq opacity} is different from that of \cite{Metzger&Pejcha17}. Our formula is more accurate than theirs when the recombination occurs at low densities $\rho\lesssim10^{-11}{\,\rm g\,cm^{-3}}$. We also adopted normalization of the Kramers opacity three times larger than that used in \citet{Metzger&Pejcha17}, to obtain better agreement with the OPAL calculations. 

The total light curve is obtained by summing the luminosity contributions from each mass/velocity shell \citep[see][]{Metzger+21}.
For a given mass profile $M_{\rm ej}(>v)$ of the form discussed below, we divide the ejecta into thin shells of velocity from $v$ to $v+dv$, which gives the shell width $dR=dvt$.
The mass contained in each shell is given by $dM=(dM_{\rm ej}/dv)dv$ and hence its density is given by $\rho=dM/(4\pi R^2 t dv)$.  With these quantities one can calculate the one-zone light curve for each shell $dL(t;v)$ and the total light curve by summing over all such shells $L(t)=\int dL(t;v)$.

The ejecta velocity distribution depends on the physical mechanism responsible for unbinding mass in stellar mergers and is uncertain theoretically.  We consider two general functional forms (exponential and power-law) for the distribution of ejecta mass above a given velocity,
\begin{align}
&M_{\rm ej}(>v)=M_{\rm ej}\begin{cases}
e^{-(v/v_{\rm ej})^{\beta_{\rm ex}}}&; \text{exponential}\ ,\\
(v/v_{\rm ej})^{-\beta_{\rm pl}}&; \text{power-law}\ ,
\end{cases}
\label{eq:Mejprofile}
\end{align}
where $\beta_{\rm ex}$, $\beta_{\rm pl}$,  and $v_{\rm ej}$ are parameters and the velocity range is $v_{\rm min}<v<v_{\rm max}$ ($v_{\rm min}=v_{\rm ej}$ for the power-law case).  In what follows it will prove useful to introduce a mean ejecta velocity $\bar{v}_{\rm E}$, which is defined by $M_{\rm ej}\bar{v}_{\rm E}^2/2=\int v^2dM/2$ and related to $v_{\rm ej}$ entering Eq.~\eqref{eq:Mejprofile} according to
\begin{align}
\frac{\bar{v}_{\rm E}}{v_{\rm ej}} =\sqrt{\frac{\int_{v_{\rm min}}^{v_{\rm max}}v^2dM}{v_{\rm ej}^2M_{\rm ej}}}\simeq\begin{cases}
\Gamma^{1/2}\left(\frac{\beta_{\rm ex}+2}{\beta_{\rm ex}}\right)&; \text{exponential}\ ,\\
\left(\frac{\beta_{\rm pl}}{\beta_{\rm pl}-2}\right)^{1/2}&; \text{power-law}\ ,
\end{cases}
    \label{eq:ve}
\end{align}
where $\Gamma(x)$ is the gamma function and we have assumed $v_{\rm min}\ll v_{\rm ej}\ll v_{\rm max}$ in the exponential case, and $\beta_{\rm pl} > 2$ and $v_{\rm max} \gg v_{\rm min}$ in the power-law case. 

To summarize, a given light curve model is defined by three main parameters: total ejecta mass $M_{\rm ej}$, initial (ejection) radius $R_0$, and mean ejecta velocity $\bar{v}_{\rm E}$ (equivalently, total kinetic energy $E_{\rm K} = M_{\rm ej}\bar{v}_{\rm E}^{2}/2$), which is related to 
$v_{\rm ej}$ through Eq.~\eqref{eq:ve}.  In what follows, we will sometimes scale the ejecta mass to a fixed fraction $\eta \equiv M_{\rm ej}/M_{\star}$ of the mass of the donor star $M_{\star}$.  We shall adopt a fiducial value of $\eta \simeq 0.25$.  This is somewhat larger than $\eta \simeq 0.1$ found in the common-envelope simulations by \citet{MacLeod+18} for a fixed accretor to donor mass ratio, 1/3.  The envelope ejection efficiency $\eta$ may also be even smaller $\lesssim 0.1$, in the case of main sequence stellar mergers (e.g., \citealt{Lombardi+02,Glebbeek+13}) or mergers of high mass-ratio binaries (e.g., \citealt{Blagorodnova+21}).

We likewise will frequently scale the mean ejecta velocity $\bar{v}_{\rm E}$ to the escape velocity of the donor at the outflow launching radius, 
\begin{equation} v_{\rm esc} = \left(\frac{2GM_{\star}}{R_{0}}\right)^{1/2} \simeq 340\,{\rm km\,s^{-1}\,} \left(\frac{M_{\star}}{3\,\Msun}\right)^{1/2}\left(\frac{R_0}{10\,\Rsun}\right)^{-1/2}\ , \label{eq:vesc}
\end{equation} 
where $G$ is the gravitational constant.
If $R_0$ equals the radius of the donor star $R_\star$, then $v_{\rm esc} = v_{\rm esc,\star}$ is the surface escape speed.  However, for $R_0 \lesssim R_{\star}$ then $v_{\rm esc} \gtrsim v_{\rm esc,\star}.$  Ejecta speeds spanning $\bar{v}_{\rm E} \sim 0.3-3\,v_{\rm esc,\star}$ are supported by LRN observations (Fig.~\ref{fig:progenitor}).

The choice of velocity distribution (Eq.~\ref{eq:Mejprofile}) introduces additional parameters, $\{\beta_{\rm ex}(\beta_{\rm pl}),v_{\rm min},v_{\rm max}\}$.  Typically, we set $v_{\rm min} = 0.1\,\bar{v}_{\rm E}$ and $v_{\rm max}= 3\,\bar{v}_{\rm E}$ but the light curve is not sensitive to these choices as long as $v_{\rm min} \ll \bar{v}_{\rm E} \ll v_{\rm max}$.  The calculation also depends on the assumed composition of the ejecta through the opacity; we assume solar metallicity ($X=0.74$ and $Z=0.02$) throughout this work, motivated by the moderately-evolved donor stars of observed LRNe progenitor binaries (Fig.~\ref{fig:progenitor}).

\section{Results}
\label{sec:results}

\subsection{Example Light Curves}
\label{sec:example}

\begin{figure}
\begin{center}
\includegraphics[width=85mm, angle=0,bb=0 0 283 221]{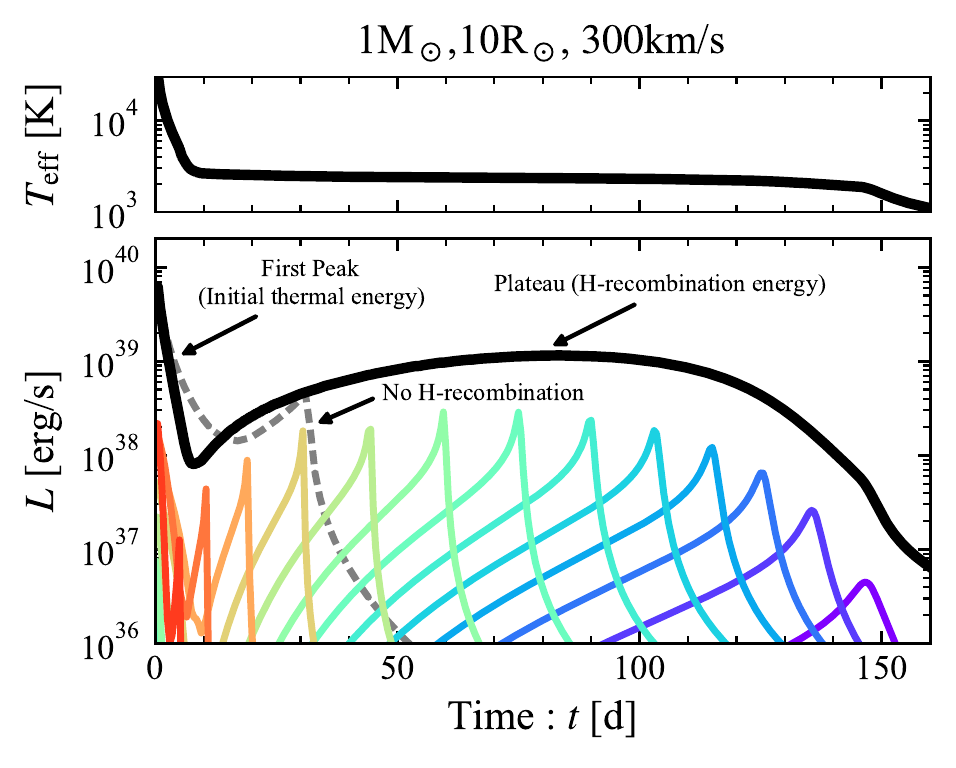}
\caption{
Example light curve for fiducial parameters: $M_{\rm ej}=\Msun$, $R_0=10\,\Rsun$, $\bar{v}_{\rm E}=300{\,\rm km\,s^{-1}}$ ($v_{\rm ej}\simeq320\,\rm km\,s^{-1}$). Colored curves denote one-zone light curves for shells with $v=50$ to $650{\,\rm km\,s^{-1}}$ (violet to orange) with a shell width $dv=5\,\rm km\,s^{-1}$.  The total light curve is characterized by two peaks, early cooling peak ($\lesssim10\,\rm d$) and late recombination plateau, produced by distinct physical processes. The gray dashed curve shows an otherwise equivalent model but neglecting energy release from recombination.  The approximate effective temperature of the emission is shown in the top panel (Eq.~\ref{eq:Teff}).   
}
\label{fig:multizone_example}
\end{center}
\end{figure}

Figure~\ref{fig:multizone_example} shows an example light curve, calculated for a total ejecta mass $M_{\rm ej} = 1\,\Msun$, initial radius $R_0 = 10\,\Rsun$, and velocity $\bar{v}_{\rm E} = 300$ km s$^{-1}$.  We employ the exponential velocity distribution, $M_{\rm ej}(>v)\propto e^{-(v/v_{\rm ej})^3}$ (Eq.~\ref{eq:ve}).  A black line shows the total light curve, while the colored lines show the separate contributions from distinct mass shells (from $v = 50$ km s$^{-1}$ to $v = 650$ km s$^{-1}$ with an interval of $50\,\rm km\,s^{-1}$ and shell width of $dv = 5$ km s$^{-1}$).

The light curve is characterized by an initial peak of high luminosity and short duration $\lesssim 1$ week followed by longer duration plateau lasting several months.  Such double-peaked light curve behavior is common in LRNe (\citealt{MacLeod+17,Metzger&Pejcha17,Pastorello+21}; Fig.~\ref{fig:obs}).  As we now describe, these two components have distinct physical origins.

\begin{figure*}
\begin{center}
\includegraphics[width=185mm,angle=0,bb=0 0 571 217]{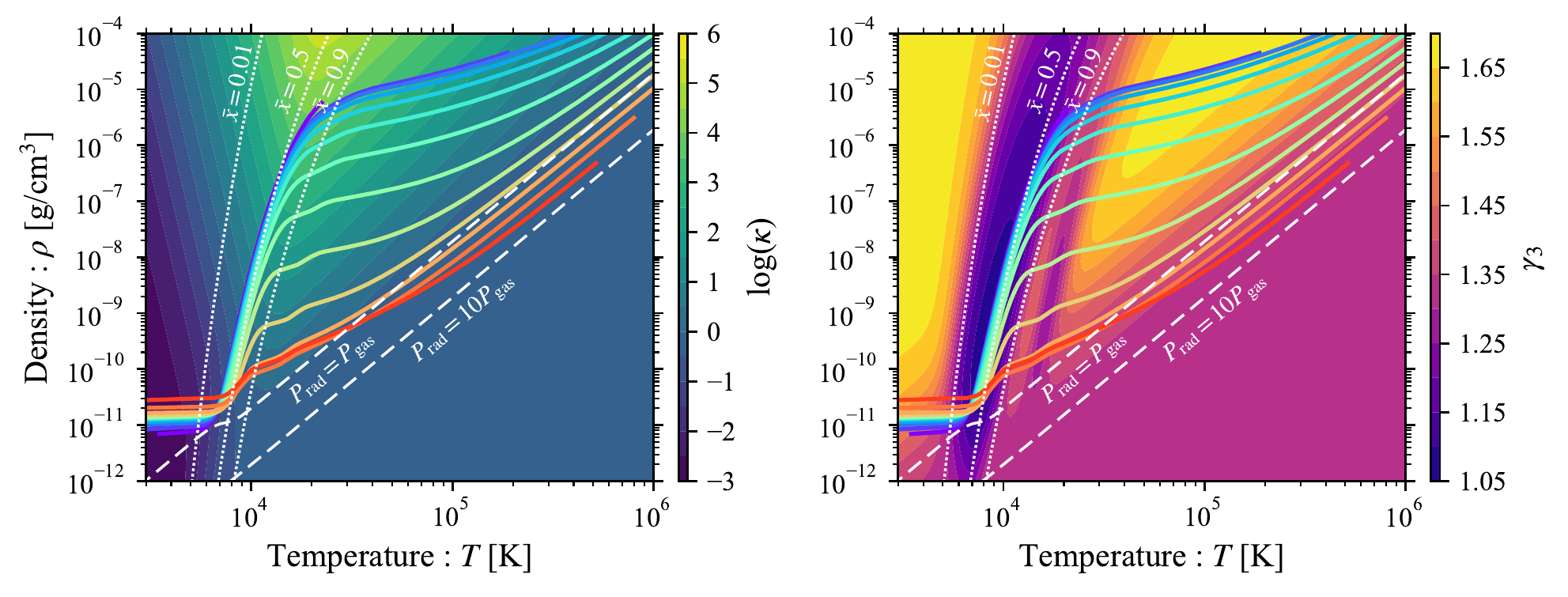}
\caption{
Colored lines show trajectories in the density-temperature plane of mass shells with $v = 50$ to $650\,\rm km\,s^{-1}$ from our fiducial model (following the same labeling scheme as in Fig.~\ref{fig:multizone_example}).  Background contours indicate the opacity $\kappa$ (left panel) and effective adiabatic index $\gamma_3$ (right panel).  Dotted and dashed contours indicate different hydrogen ionization fractions $\bar{x}$ and ratios of gas-to-radiation pressure.  Prior to hydrogen recombination, each shell decompresses and cools adiabatically with $\gamma_3 \in [4/3,5/3],$ moving from the upper right hand corner of the plot to the lower left.  However, once recombination starts at $T \simeq 10^{4}$ K, the energy released from recombination holds the temperature roughly constant at $T \simeq 10^{4}$ K ($\gamma_3 \simeq 1.1$), causing a convergence of the mass shell trajectories.  However, as the density continues to drop, once $P_{\rm rad} > P_{\rm gas}$ ($\rho \lesssim 10^{-11}$ g cm$^{-3}$ for $T \sim T_{\rm i} \simeq 5000$ K), $\gamma_3$ again starts to rise, causing $T$ to decrease faster and recombination to complete.  The resulting decrease in the opacity leads to the light curve peaking, thus motivating our analytic estimate of the transient duration (Eq.~\ref{eq:tpl}). }
\label{fig:rhoT_trajectories}
\end{center}
\end{figure*}

\begin{figure}
\begin{center}
\includegraphics[width=85mm,angle=0,bb=0 0 268 206]{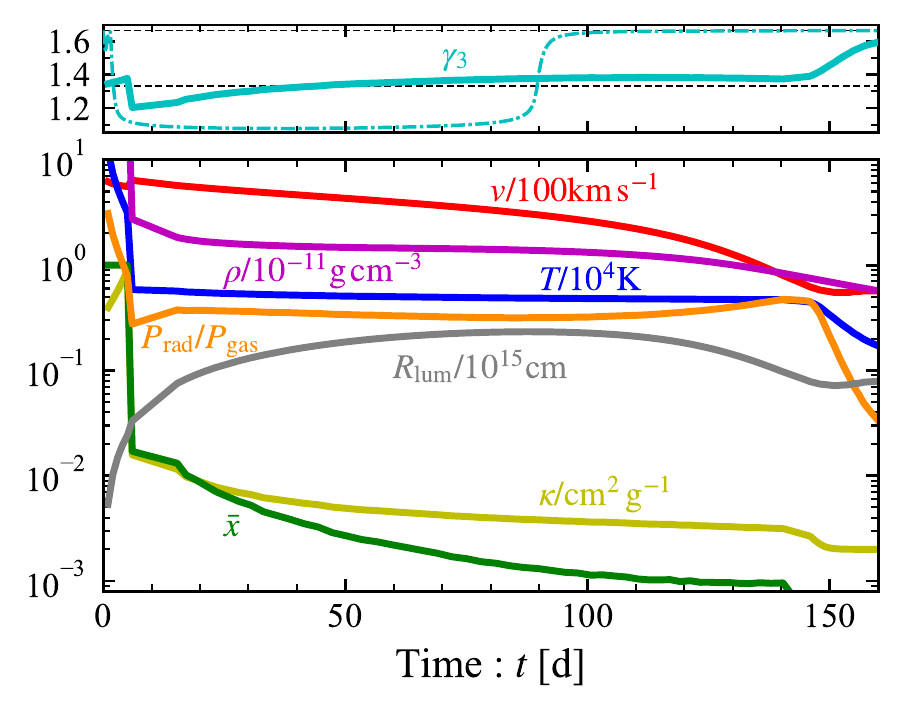}
\caption{
Time evolution of key quantities as measured at the shell which dominates the luminosity at a given epoch, for the fiducial model shown in Fig.~\ref{fig:multizone_example}. The top panel shows the effective adiabatic index $\gamma_3$ (Eq.~\ref{eq:gamma3}); for comparison dashed lines show the usual values for radiation- (gas-) pressure-dominated gas, $\gamma_3=4/3$ ($5/3$), absent recombination effects. A dash-dotted curve in the top panel shows the value of $\gamma_3$ for the single fixed shell with $v=\bar{v}_{\rm E}$.}
\label{fig:fiducial_peakquantities}
\end{center}
\end{figure}

\subsubsection{Early Cooling Peak}

At early stages of evolution, the fastest, lower-mass layers of the ejecta dominate the transient luminosity.  As we will confirm below, these layers are typically dominated by radiation pressure and their energy source is ``cooling'' emission produced by the initially hot ejecta (e.g., \citealt{MacLeod+17}), similar to the mechanism responsible for powering the plateau of SNe II \citep{Arnett80}.  The duration and luminosity of this early emission phase are given by \citep[e.g.,][]{Arnett80,Padmanabhan00}
\begin{eqnarray}
t_{\rm pk} \simeq \biggl(\frac{\kappa_{\rm es} M}{4\pi c v}\biggl)^{1/2}\simeq 6.7{\,\rm d\,}\biggl(\frac{M}{10^{-2}\,\Msun}\biggl)^{1/2}\biggl(\frac{v}{500\,\rm km\,s^{-1}}\biggl)^{-1/2}\ ,
	\label{eq:tpk}
\end{eqnarray}
\begin{eqnarray}
L_{\rm pk} &\simeq& \frac{\frac{1}{2}Mv^2(R_0/R_{\rm pk})}{t_{\rm pk}} \simeq 1.0\times10^{39}{\,\rm erg\,s^{-1}}\times \nonumber \\
&& \biggl(\frac{R_0}{10\,\Rsun}\biggl)\biggl(\frac{v}{500\,\rm km\,s^{-1}}\biggl)^2\ ,
	\label{eq:Lpk}
\end{eqnarray}
\begin{eqnarray}
E_{\rm pk} &\simeq& L_{\rm pk}t_{\rm pk} \simeq 6 \times 10^{44}{\,\rm erg\,} \times \\ \nonumber 
& & \left(\frac{M}{10^{-2}\,\Msun}\right)^{1/2}\left(\frac{R_0}{10\,\Rsun}\right)\left(\frac{v}{500\rm \,km\,s^{-1}}\right)^{3/2},
	\label{eq:Epk}
\end{eqnarray}
where $M$ is the ejecta mass of the fastest layer, $R_{\rm pk}=vt_{\rm pk}$, and we have assumed that electron scattering of ionized hydrogen $\kappa_{\rm es} \simeq 0.32$ cm$^{2}$ g$^{-1}$ dominates the opacity during the early peak.

The assumptions entering Eqs.~(\ref{eq:tpk})-(\ref{eq:Epk}) can now be checked. 
The ratio of radiation to gas pressure, on the timescale $\sim t_{\rm pk}$, can be estimated as
\begin{align}
\frac{P_{\rm rad}}{P_{\rm gas}}\biggl|_{t_{\rm pk}} &\simeq5.1\left(\frac{M}{10^{-2}\Msun}\right)^{-1/4}\left(\frac{R_0}{10\,\Rsun}\right)^{3/4}\left(\frac{v}{500\,{\rm km\,s^{-1}}}\right)^{3/2}
    \nonumber\\
&\simeq0.98\,\eta_{0.25}^{-1/4}\left(\frac{v}{v_{\rm esc}}\right)^{3/2}\left(\frac{M_{\star}}{3\,\Msun}\right)^{1/2},
\label{eq:PradPgas}
\end{align}
where (assuming $P_{\rm rad} \gg P_{\rm gas}$) we have used $Mv^{2}/2 = (4\pi/3)R_0^{3} aT_0^{4}$ to derive the initial temperature and $T(t_{\rm pk})=T_0(R_0/R_{\rm pk})$.
In the second line of Eq.~(\ref{eq:PradPgas}) we have set $M=M_{\rm ej}$ and scaled the ejecta mass to donor star mass $M_{\star}$ via $\eta = M_{\rm ej}/M_{\star}$ ($\eta_{0.25} = \eta/0.25$) and the ejecta velocity to the escape speed from the outflow launching radius (Eq.~\ref{eq:vesc}).  Note that as long as the radiation pressure dominates initially, the pressure ratio does not evolve significantly until the light curve peaks.\footnote{Neglecting the ionization energy and radiation energy loss, we can solve Eqs.~\eqref{eq:internal_energy}, \eqref{eq:evolution}, and \eqref{eq:gamma3} to obtain the time evolution of the pressure ratio ($\alpha = P_{\rm rad}/P_{\rm gas}$, see Eq.~\ref{eq:PradPgas}): 
\begin{align}
\frac{t}{t_0}=\biggl(\frac{\alpha}{\alpha_0}\biggl)^{-1/3}\biggl(\frac{2\alpha+1}{2\alpha_0+1}\biggl)^{3}e^{\frac{32}{3}(\alpha_0-\alpha)}\ ,
    \nonumber
\end{align}
where $\alpha_0$ is the initial ratio at $t_0=R_0/v$.  We see that if the radiation pressure dominates even a little over gas pressure, $\alpha_0\gtrsim$ a few, the pressure ratio does not evolve until radiative cooling becomes important. This is because the radiation entropy, which is conserved under adiabatic expansion, is proportional to the pressure ratio.}
This allows us to estimate the minimum velocity of the radiation-pressure dominated shell, $v\simeq500\,\rm km\,s^{-1}$ for our fiducial model with $M\propto v^3 e^{-(v/v_{\rm ej})^3}$, consistent with Fig.~\ref{fig:multizone_example}.

We see that the radiation pressure will dominate in the high-velocity tail of the ejecta $v \gg v_{\rm esc}$ and that Eqs.~(\ref{eq:tpk})-(\ref{eq:Epk}) indeed determine the early-time peak emission from these layers (and hence the early-peak of the transient as a whole).  However, for the bulk of the ejecta for which $v \lesssim v_{\rm esc}$, gas pressure will dominate and the early cooling emission from these layers will be reduced compared to $L_{\rm pk}$ in Eq.~(\ref{eq:Lpk}) because of the greater adiabatic losses experienced by the ejecta for $\gamma = 5/3$ ($T\propto R^{-2}$) than $\gamma = 4/3$.  Previous models for LRNe employ analytic formula for the luminosity and duration (e.g., \citealt{Popov93}) which implicitly assume radiation pressure dominates, as is justified in SNe due to their higher ejecta velocities and specific internal energy.

The ratio of Kramers to electron scattering opacity during the early peak can also be checked:
\begin{eqnarray}
\frac{\kappa_{\rm es}}{\kappa_{\rm K}}\biggl|_{t_{\rm pk}} &\simeq&
5.3 \left(\frac{M}{10^{-2}\,\Msun}\right)^{-3/8}\left(\frac{R_0}{10\,\Rsun}\right)^{7/8}\left(\frac{v}{500\,{\rm km\,s^{-1}}}\right)^{3/2}\ ,
\end{eqnarray}
where again we take $\kappa_{\rm es} \simeq 0.32$ cm$^{2}$ g$^{-1}$.  This confirms that $\kappa_{\rm es} \gtrsim \kappa_{\rm K}$, as assumed in Eqs.~(\ref{eq:tpk})-(\ref{eq:Epk}).

\subsubsection{Late Recombination Plateau}

The longer plateau-like emission phase in Fig.~\ref{fig:multizone_example} is not powered by the ejecta's initial thermal energy, but instead by hydrogen recombination.  After forming its initial luminosity peak, the ejecta cool down and hydrogen starts to recombine at temperatures $T\sim10^{4}\,\rm K$.  Once recombination sets in, the additional energy supply causes the adiabatic index to decrease to $\gamma_3-1\simeq0.1$, thus slowing the temperature evolution greatly compared to that of ordinary adiabatic expansion absent recombination $(\gamma_3 \lesssim 5/3)$.  By the time a given shell completes recombination at temperatures $T_{\rm i}\simeq4000-6000\,\rm K$ most of the radiation energy has been emitted, generating a second peak in its light curve.  The sum of the second peak from all mass shells powers the $\sim 100$ day plateau in Fig.~\ref{fig:multizone_example}.

The duration of the plateau is roughly given by the timescale over which the recombination completes at $T \gtrsim T_{\rm i}$.  
We find that recombination completes at roughly a constant characteristic density $\rho_{\rm i}\sim10^{-11}\,\rm g\,cm^{-3}$.  As explained in Fig.~\ref{fig:rhoT_trajectories}, this density corresponds to roughly where $P_{\rm rad} \sim P_{\rm gas}$ for $T \sim T_{\rm i}$.  The increasing adiabatic index once $P_{\rm rad} \gtrsim P_{\rm gas}$ triggers the completion of ionization of each mass shell, causing the opacity to decrease and leading to the light curve peak.

The duration of the recombination plateau is thus set by the timescale for the density to reach $\rho_{\rm i}$, 
\begin{align}
t_{\rm pl}\sim\biggl(\frac{3M_{\rm ej}}{4\pi\rho_{\rm i}\bar{v}_{\rm E}^3}\biggl)^{1/3}\simeq140{\,\rm d\,} \rho_{\rm i,-11}^{-1/3}\biggl(\frac{M_{\rm ej}}{\Msun}\biggl)^{1/3}\biggl(\frac{\bar{v}_{\rm E}}{300\,{\rm km\,s^{-1}}}\biggl)^{-1}\ ,
	\label{eq:tpl}
\end{align}
where $\rho_{\rm i,-11}=\rho_{\rm i}/10^{-11}\,\rm g\,cm^{-3}$.
Since the plateau is powered by the hydrogen recombination ($\varepsilon_{\rm H} = 13.6$ eV per hydrogen atom), the radiated energy and luminosity can be written 
\begin{eqnarray}
E_{\rm pl} &\sim& f_{\rm ad}\frac{\varepsilon_{\rm H}XM_{\rm ej}}{m_{\rm p}}\simeq 5.8\times10^{45}{\,\rm erg\,}f_{\rm ad,0.3}\biggl(\frac{M_{\rm ej}}{\Msun}\biggl)\ ,
    \label{eq:Epl}
    \end{eqnarray}
    \begin{eqnarray}
L_{\rm pl} &\sim& \frac{E_{\rm pl}}{t_{\rm pl}}\simeq 4.8\times10^{38}{\,\rm erg\,s^{-1}\,}f_{\rm ad,0.3}\rho_{\rm i,-11}^{1/3}\times \nonumber \\ && \biggl(\frac{M_{\rm ej}}{\Msun}\biggl)^{2/3}\biggl(\frac{\bar{v}_{\rm E}}{300\,{\rm km\,s^{-1}}}\biggl)\     \label{eq:Lpl},
\end{eqnarray}
where the dimensionless parameter $f_{\rm ad} = 0.3f_{\rm ad,0.3} < 1$ accounts for inefficiency in radiating the recombination energy due to adiabatic losses.  Helium recombination also contributes, but since it occurs at higher temperature (at hence much higher optical depth), its effect on the light curve is small compared to that of hydrogen recombination.  

In Appendix \ref{sec:appendix} we present a one-zone toy light curve model for a hydrogen-rich explosion, in the spirit of \citet{Popov93} but assuming gas-pressure instead of radiation dominates and accounting for recombination energy through an (assumed constant) effective value of the adiabatic index $\gamma_3$; for values of $\gamma_3 \sim 1$ expected during recombination the analytic expressions for the plateau luminosity (Eq.~\ref{eq:appen Lpl}), duration (Eq.~\ref{eq:appen tpl}), and radiated energy (Eq.~\ref{eq:appen Epl}), are quite similar to those  Eqs.~\eqref{eq:tpl}-\eqref{eq:Lpl} above.
This is remarkable because the conceptual motivation of the two derivations are different: the plateau duration derived in Appendix \ref{sec:appendix} (Eq.~\ref{eq:appen tpl}) is dictated by the photon diffusion timescale through the recombining ejecta shell, while Eq.~\eqref{eq:tpl} was derived assuming recombination occurs at roughly fixed density.

The top panel of Fig.~\ref{fig:multizone_example} shows the evolution of the effective temperature, which we have defined by
\begin{align}
T_{\rm eff}\equiv \biggl(\frac{L}{4\pi \sigma_{\rm SB} R_{\rm lum}^2}\biggl)^{1/4}\ ,
    \label{eq:Teff}
\end{align}
where $\sigma_{\rm SB}$ is the Stefan-Boltzmann constant and $R_{\rm lum}$ is the radius of the shell contributing the majority of the luminosity at any epoch.  During the initial peak, $T_{\rm eff} \simeq10^4\,\rm K$ before cooling during the recombination plateau to an almost fixed valued $T_{\rm eff} \simeq3000-4000\,\rm K$, consistent with the ``red'' colors of LRNe.  This temperature stability occurs for the same reason as in SNe II, namely that the recombining shell recedes monotonically in the velocity coordinate but the expansion effect compensates and leads to the almost spatially fixed location of the shell (see Fig.~\ref{fig:fiducial_peakquantities}).  We emphasize however, that$-$unlike in SNe$-$the plateau is powered not by radioactivity or the thermal energy released from the initial explosion, but rather from the recombination energy itself. We confirm this by showing an otherwise identical light curve model but calculated with recombination energy artificially excluded, as a gray line in Fig. \ref{fig:multizone_example}, demonstrating the absence in this case of an extended plateau phase.  This further implies that standard relationships applied to model SN light curves (e.g., \citealt{Popov93}) should not be applied to LRNe.

Figure \ref{fig:fiducial_peakquantities} shows the time evolution of key quantities for the fiducial model, as measured at the shell dominating the luminosity at each epoch ($R_{\rm lum}$).  As expected, $P_{\rm rad} \gg P_{\rm gas}$ during the early peak when the highest velocity layers dominate the emission, while $P_{\rm rad} \lesssim P_{\rm gas}$ during the later recombination peak. Also note the mean ionization fraction $\bar{x}$ is low $\lesssim 10^{-2}$ during the prolonged recombination plateau, and that electron scattering opacity $\kappa_{\rm es} \simeq 0.32\,\bar{x}$ cm$^{2}$ g$^{-1}$ becomes comparable to H$^-$ opacity. During the recombination phase, the effective adiabatic index $\gamma_3 \lesssim 4/3$ due to the energy released by recombination offsetting losses due to adiabatic expansion. Note that although the index $\gamma_3$ of $R_{\rm lum}$ is close to $4/3$, this is not related to radiation pressure. Rather, during the brief epoch when a shell dominates the total luminosity, its adiabatic index quickly increases from $\gamma_3\simeq1.1$ to $5/3$ passing through an value coincidentally close to $4/3$ (as illustrated by a dash-dotted curve in Fig.~\ref{fig:fiducial_peakquantities}).

\subsubsection{Variations About the Fiducial Model}

\begin{figure*}
\begin{center}
\includegraphics[width=160mm, angle=0,bb=0 0 508 224]{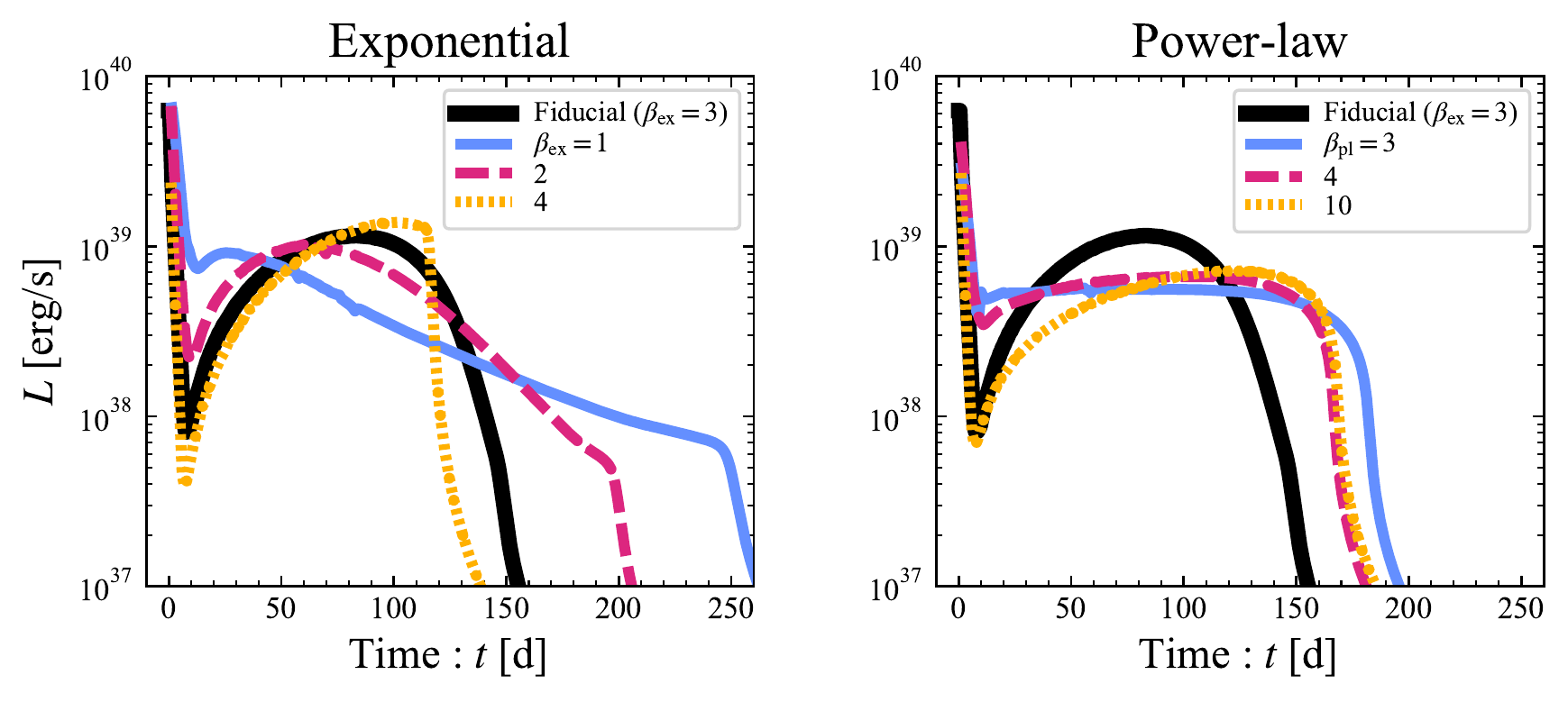}
\caption{Impact on the fiducial model light curve (Fig.~\ref{fig:multizone_example}) of changing the shape of the ejecta velocity distribution (Eq.~\ref{eq:Mejprofile}), for a fixed mean velocity $\bar{v}_{\rm E}$.  The left panel shows the effect of varying the parameter $\beta_{\rm ex}$ which enters the exponential velocity distribution $M_{\rm ej}(>v) \propto e^{-(v/v_{\rm ej})^{\beta_{\rm ex}}}$, while the right panel shows the impact of varying the power-law index $\beta_{\rm pl}$ for $M_{\rm ej}(>v) \propto v^{-\beta_{\rm pl}}$.  Although the light curve shape changes for different velocity distributions, the qualitative features of an early peak followed by a second peak or plateau are robust to the details.}
\label{fig:changing_velocityprofile}
\end{center}
\end{figure*}

Figure~\ref{fig:changing_velocityprofile} shows light curve models calculated for ejecta properties otherwise identical to the fiducial model (exponential model with $\beta_{\rm ex} = 3$) but varying the shape of the velocity distribution at fixed $\bar{v}_{\rm E}$ (Eq.~\ref{eq:Mejprofile}).  For the exponential velocity profile, the duration of the plateau shortens with increasing values of $\beta_{\rm ex}$, but the qualitative shape does not change significantly unless $\beta_{\rm ex} \lesssim 2$.  Likewise, in the case of a power-law profile, the light curve shape is not qualitatively dependent on the power-law index $\beta_{\rm pl}$.  Motivated thus, we hereafter fix the velocity profile to the fiducial exponential $\beta_{\rm ex} = 3$.

\begin{figure}
\begin{center}
\includegraphics[width=85mm,angle=0,bb=0 0 275 211]{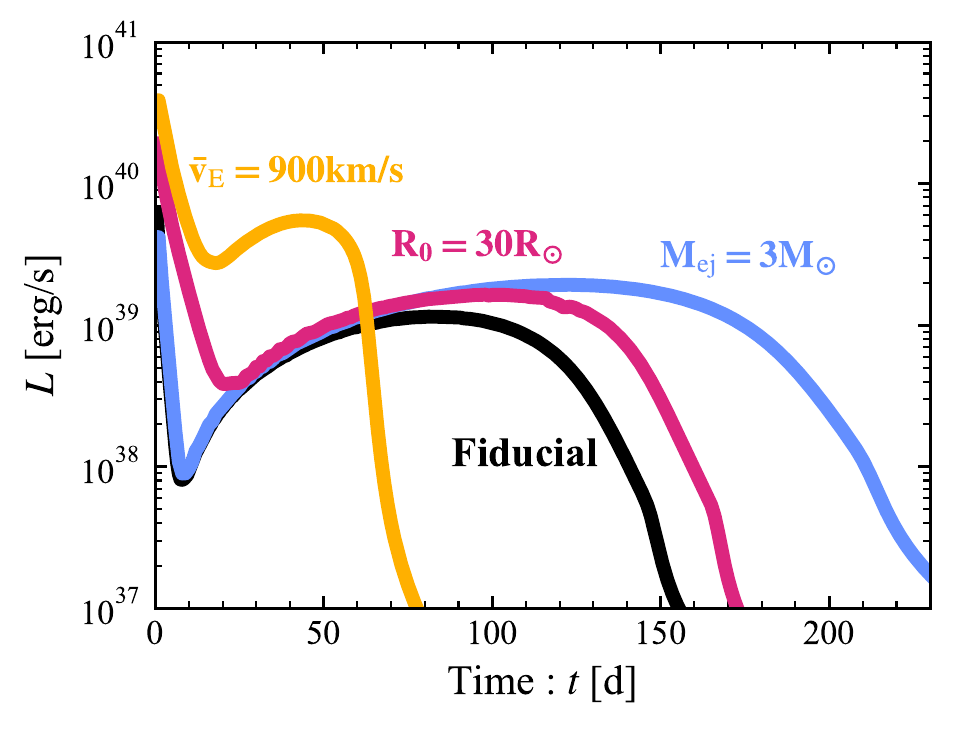}
\caption{Comparison of the light curve for the fiducial model ($M_{\rm ej} = 1\,\Msun$; $R_0 = 10\,\Rsun$; $\bar{v}_{\rm E} = 300$ km s$^{-1}$; Fig.~\ref{fig:multizone_example}) to otherwise similar models in which: the ejecta mass is increased to $3\,\Msun$ (blue); the initial radius is increased to $R_0 = 30\,\Rsun$ (red); the mean velocity is increased to $\bar{v}_{\rm E} =900$ km s$^{-1}$ (orange).  The trends in the luminosity and duration of the first and second peaks follow Eqs.~\eqref{eq:tpk}, \eqref{eq:Lpk} and Eqs.~\eqref{eq:tpl}, \eqref{eq:Lpl}, respectively. 
}
\label{fig:lc_varyparameters}
\end{center}
\end{figure}

Figure~\ref{fig:lc_varyparameters} compares the light curve for the fiducial model ($M_{\rm ej} = 1\,\Msun$; $R_0 = 10\,\Rsun$; $\bar{v}_{\rm E} =300$ km s$^{-1}$) to otherwise similar models but for which $\{M_{\rm ej}, R_0, \bar{v}_{\rm E}\}$ are tripled from their fiducial values.  As expected from Eqs.~(\ref{eq:tpk})-(\ref{eq:Epk}), larger values of $R_0$ and $\bar{v}_{\rm E}$ increase the luminosity of the early peak.  As expected from Eqs.~\eqref{eq:tpl}-\eqref{eq:Lpl}, the properties of the recombination plateau are roughly independent of $R_0$.  Likewise, the plateau gets shorter but more luminous with increasing velocity, while both the duration and luminosity increase for larger ejecta mass.  

As can be seen in Fig.~\ref{fig:obs}, many LRNe light curves exhibit multi-peak light curves in which the first peak is substantially brighter than the longer plateau to follow (e.g., V838 Mon: \citealt{Munari+02}; M31-OT2011: \citealt{Kurtenkov+15,MacLeod+17}; AT2019zhd: \citealt{Pastorello+21}; AT2020kog: \citealt{Pastorello+21b}) and others in which a distinct first peak is more subtle or non-existent (e.g., V1309 Sco: \citealt{Mason+10,Tylenda+11}; AT2020hat: \citealt{Pastorello+21b}).  This difference may indicate the presence or absence of a high-velocity ejecta component or different progenitor's properties. To understand the latter effect, using Eqs.~(\ref{eq:Lpk}) and (\ref{eq:Lpl}), we see that the ratio of the cooling envelope peak luminosity to that of the recombination plateau can be estimated as
\begin{eqnarray}
\frac{L_{\rm pk}}{L_{\rm pl}} &\sim& 3\,f_{\rm ad,0.3}^{-1}\left(\frac{M_{\rm ej}}{\Msun}\right)^{-2/3}\left(\frac{R_0}{10\,\Rsun}\right)\left(\frac{\bar{v}_{\rm E}}{300\,{\rm km\,s^{-1}}}\right) \nonumber \\
&\sim& 4\,f_{\rm ad,0.3}^{-1}\eta_{0.25}^{-2/3}\left(\frac{M_{\star}}{3\,\Msun}\right)^{-1/6}\left(\frac{R_0}{10\,\Rsun}\right)^{1/2}\left(\frac{\bar{v}_{\rm E}}{v_{\rm esc}}\right)\ ,
    \label{eq:luminosity_ratio}
\end{eqnarray}
where we have taken $\rho_{\rm i,-11} = 1$ and assumed that the outer layers of the ejecta responsible for powering the first peak have a velocity twice as large as that of the bulk, $v\simeq 2\,\bar{v}_{\rm E}$.  Thus, a range of values $L_{\rm pk}/L_{\rm pl} \sim 1-10$ can be obtained.

\subsection{Parameter Study of Stellar Mergers Across the HR Diagram}
\label{sec:parameterstudy}

We now explore the light curve behavior across a larger parameter space of ejecta mass $M_{\rm ej} \sim 10^{-3}-10\,\Msun$ and initial radius $R_0 \sim 10^{-2}-100\,\Rsun$ covering those expected from a wide range of stellar and even sub-stellar merger events.
For each model we set the mean ejecta velocity to the escape speed from $R_0$, $\bar{v}_{\rm E}=v_{\rm esc}$ (Eq.~\ref{eq:vesc}), where the stellar mass is related to the ejecta mass by $M_{\rm ej}=\eta M_{\star}$ with $\eta = 0.25$.  For each light curve model, we calculate the total radiated energy $E_{\rm rad}=\int L dt$.  We also define a characteristic transient duration, $t_{90}$, as the interval over which 90\% of the total energy is radiated, $0.9\,E_{\rm rad}=\int^{t_{90}}Ldt$.  Finally, from the radiated energy and duration we define a mean luminosity $L_{90} \equiv E_{\rm rad}/t_{90}$.

\begin{figure*}
\begin{center}
\includegraphics[width=160mm, angle=0,bb=0 0 552 503]{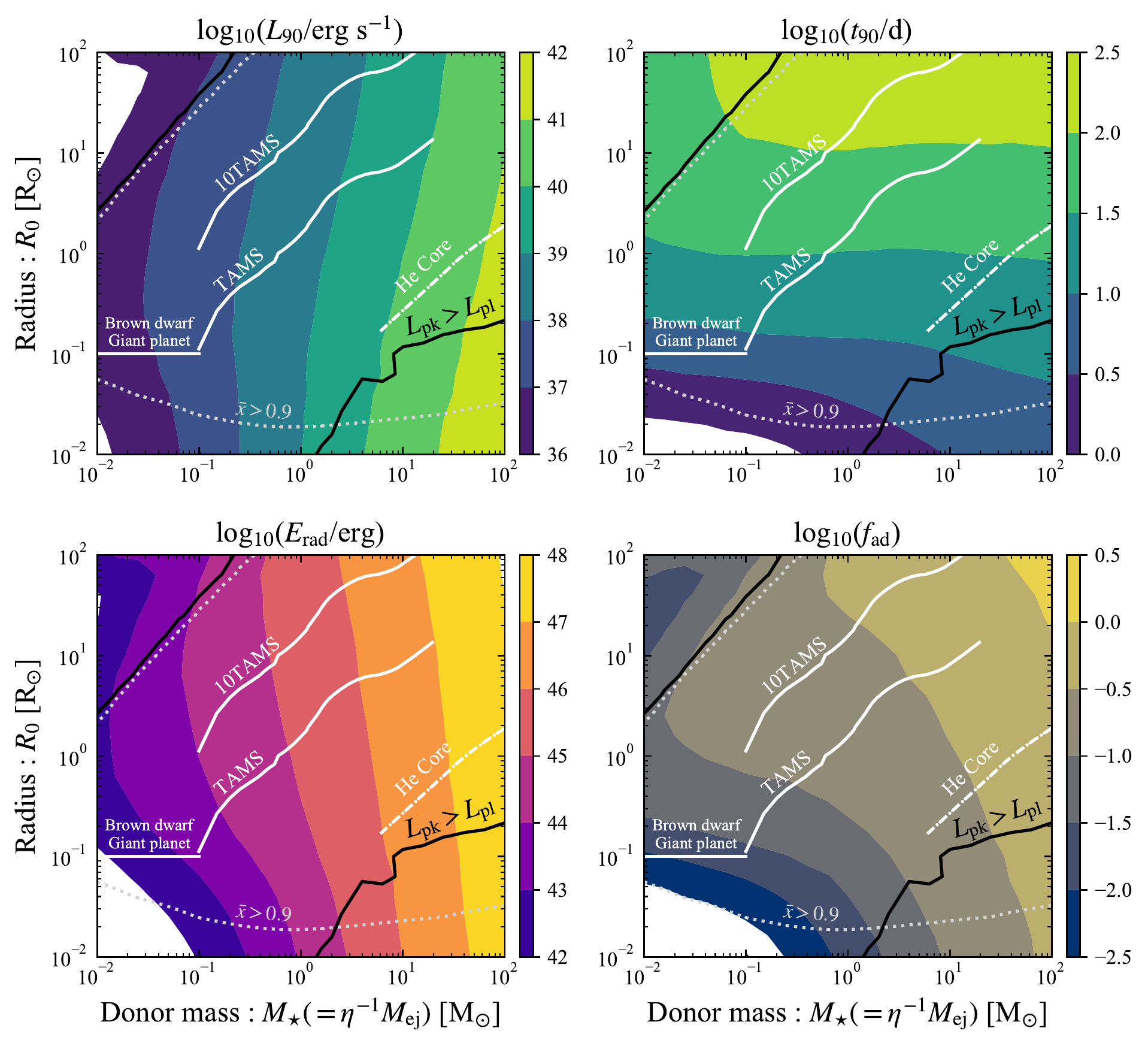}
\caption{Characteristic luminosity, timescale, radiated energy, and ratio of radiated energy to recombination energy calculated for a range of stellar (donor) mass and radius for our fiducial velocity profile (Eq. \ref{eq:Mejprofile}). Ejecta mass is linked to the stellar mass via $M_{\rm ej}=\eta M_{\star}$ ($\eta=0.25$) and velocity is set to the escape velocity.
White curves denote the mass and radius relation for TAMS, ten times larger radius of TAMS (``10TAMS''), and brown dwarf and giant planet.
A white dash-dotted curve shows the radius of He core for the corresponding ZMAS mass ($M_{\star}$), which gives a rough lower limit of the mass-ejection radius $R_0$ from an evolved star.  A black solid line denotes the contour along which the peak luminosities of the initial cooling phase and later recombination plateau are equal (see Fig.~\ref{fig:multizone_example}), as calculated directly from the light curve models. 
A gray dotted curve shows the contour on which the mean ionization degree of the typical shell ($v=\bar{v}_{\rm E}$) is $\bar{x}=0.9$ (partially ionized), at the initial time.
}
\label{fig:dist1}
\end{center}
\end{figure*}

Figure~\ref{fig:dist1} shows the distribution of the characteristic luminosity, duration, and radiated energy.  We also plot the ratio of the radiation energy to the total hydrogen recombination energy $f_{\rm ad} \equiv E_{\rm rad}/(\varepsilon_{\rm H}XM_{\rm ej}/\MP$).  For comparison we show tracks of stellar radius versus terminal-age main sequence (TAMS), from the MIST database of MESA stellar evolution models  \citep{Dotter16,Choi+16,Paxton+11,Paxton+13,Paxton+15,Paxton+19}, as well as a hypothetical state of post-main sequence evolution in which the stellar radius is 10 times larger than its TAMS value (``10TAMS'').  These bracket the state of LRN progenitor stars from pre-transient imaging studies (Fig.~\ref{fig:progenitor}).  For matter ejected from the stellar surface, we expect $R_0 \sim R_{\star}$, while for matter ejected from deeper within the donor star $R_0 < R_{\star}$.

\begin{figure*}
\begin{center}
\includegraphics[width=160mm, angle=0,bb=0 0 500 229]{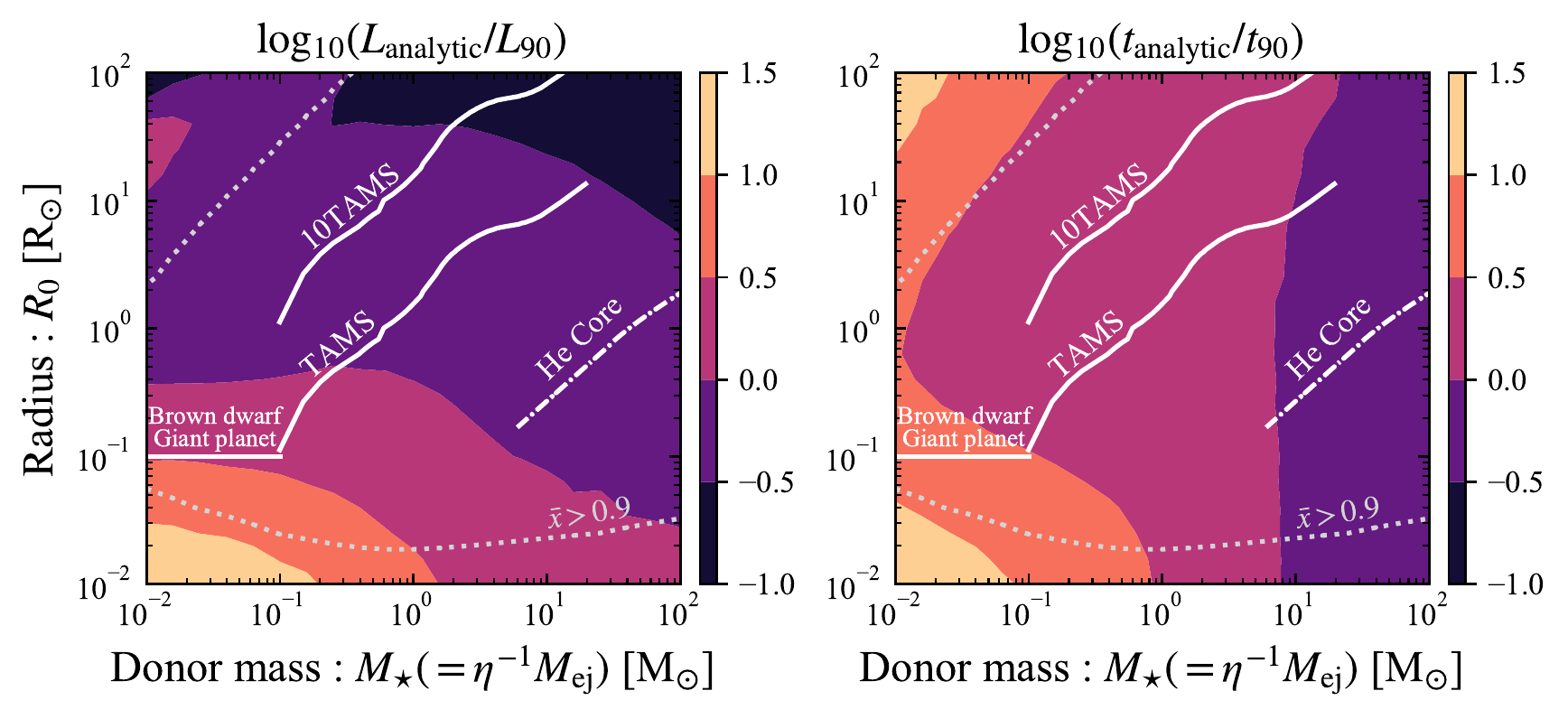}
\caption{The same parameter space shown in Fig.~\ref{fig:dist1} but showing the ratio of the plateau luminosity and duration predicted by our analytical formulae (Eqs.~\ref{eq:tpl},\ref{eq:Lpl}; for fixed $f_{\rm ad,0.3} = 1$ and $\rho_{\rm i,-11} = 1$) to those obtained from the full numerical model (Fig.~\ref{fig:dist1}). The analytic formulae accurately predict the luminosity and duration within a factor of $\lesssim 3$ along the parameter space surrounding the TAMS sequence.}
\label{fig:ratio_ourmodel}
\end{center}
\end{figure*}

Across most of the parameter space, we find that the total radiated energy is dominated by the later recombination plateau, consistent with a ``by eye'' integration of LRN light curves (Fig.~\ref{fig:obs}).
The parameter dependencies of the luminosity, duration, and radiated energy are roughly consistent with those expected from Eqs.~\eqref{eq:tpl}-\eqref{eq:Lpl}.  In particular, for $\bar{v}_{\rm E} = v_{\rm esc} \propto M_{\star}^{1/2}R_0^{-1/2}$ and $M_{\rm ej} \propto M_{\star}$, these formulae predict $t_{90} \sim t_{\rm pl} \propto M_{\star}^{1/6}R_0^{1/2}$, $L_{90} \sim L_{\rm pl} \propto M_{\star}^{7/6}R_0^{-1/2}$, and $E_{\rm rad}\sim E_{\rm pl} \propto M_{\star}$, consistent with the contours in Fig.~\ref{fig:dist1}.  
These relations no longer hold when the ejecta is already partially ionized at the initial time (upper left and bottom regions in Fig.~\ref{fig:dist1}).  Figure \ref{fig:ratio_ourmodel} shows the ratios of the transient luminosity and duration as predicted by our analytic formulae (Eqs.~\ref{eq:tpk},\ref{eq:Lpk}) to those obtained from the full numerical calculations (Fig.~\ref{fig:dist1}), demonstrating good agreement (to within a factor $\lesssim 3$) across a band encompassing the entire TAMS sequence.  

For mergers that occur when the donor is near the end of the main sequence ($R \sim R_{\rm TAMS}$) or after evolving moderately up the giant branch ($R \sim 10R_{\rm TAMS}$) the predicted luminosities are in the range $\sim 10^{39}-10^{40}$ erg s$^{-1}$, broadly consistent with the bulk of the LRN population. 
Luminosities up to $\sim 10^{41}$ erg s$^{-1}$ may be possible in extreme cases for efficient ejection of the stellar envelope $\eta\sim 1$ from deep within the potential of the stellar core ($R_0 \ll R_{\rm TAMS}$).
The minimum launching radius of the donor star envelope for an evolved star is determined by the size of the final He core, which we show as a white dash-dotted curve in Fig.~\ref{fig:dist1} (we adopt the analytic $M_{\star}-M_{\rm He}$ relationship from \citealt{Woosley19} and employ the $M_{\rm He}-R_{\star}$ relationship from Eq.~4 of \citealt{Schaerer&Maeder92}; note that $\eta \lesssim 0.5-0.9$ in this case).  Even under the most optimistic assumptions, the model is strained to reach luminosities $L >10^{41}$ erg s$^{-1}$, which can therefore be taken as a theoretical upper limit (absent other energy sources not included in our fiducial model; Section \ref{sec:shocks}).  

Typical values of the radiative efficiency of the recombination energy are $f_{\rm ad}=E_{\rm rad}/(\varepsilon_{\rm H}XM_{\rm ej}/\MP) \simeq 0.1-0.3$.  Across most of the parameter space, the luminosity of the initial thermal peak $L_{\rm pk}$ will stick out above that of the recombination plateau $L_{\rm pl}$, even though the radiated energy is dominated by the plateau.  The parameter space for which $L_{\rm pk} > L_{\rm pl}$ is delineated by a solid black line in Fig.~\ref{fig:dist1}, roughly consistent with our analytic estimate in Eq.~\eqref{eq:luminosity_ratio}.
Note however that having $L_{\rm pk} > L_{\rm pl}$ does not necessarily imply the early cooling peak will be detected, because its duration is significantly shorter than the plateau, $t_{\rm pk}/t_{\rm pl}\lesssim 0.01-0.1$, and could be missed depending on the discovery epoch and observational cadence of the survey (Fig.~\ref{fig:obs}). 

\subsection{Application to Individual LRNe}
\label{sec:IndividualLRN}

\begin{figure*}
\begin{center}
\includegraphics[width=160mm, angle=0,bb=0 0 474 256]{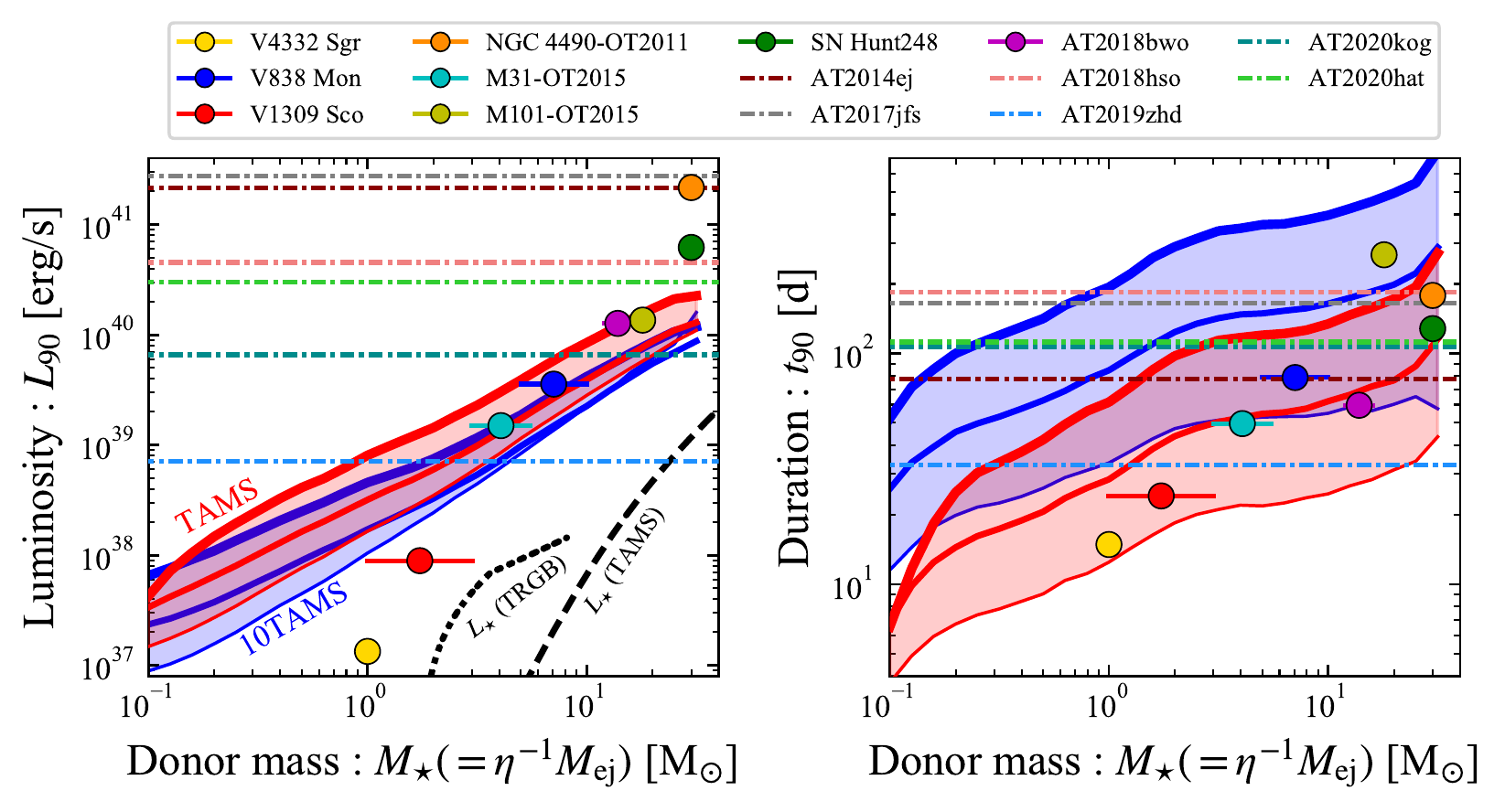}
\caption{Luminosity $L_{90}$ (left) and duration $t_{90}$ (right) of stellar merger transients compared to LRNe with progenitor donor masses $M_{\star}$ inferred from pre-transient imaging (Table \ref{table data}).  Red and blue shaded regions show the predictions of our light curve model as a function of donor mass $M_{\star}$ for a range of $\eta=M_{\rm ej}/M_{\star}= 0.01$ (thinnest lines) to $\eta = 1$ (thickest lines), and where we have assumed $\bar{v}_{\rm E} = v_{\rm esc,\star} = (2GM_{\star}/R_{\star})^{1/2}$ for stars on the TAMS or 10 times their TAMS radii.   In the left panel, a dashed curve denotes the stellar luminosity at TAMS \citep{Hurley+00}, while a dotted line shows the luminosity at the tip of the red giant branch (\citealt{Paczynski70,Cummings+18}).}
\label{fig:L90t90}
\end{center}
\end{figure*}

Figure \ref{fig:L90t90} shows the mean luminosity $L_{90}$ and duration $t_{90}$ for a sample of LRNe (Table~\ref{table data}; Fig.~\ref{fig:obs}) as a function of the donor mass $M_{\star}$ inferred from pre-transient imaging observations when available.\footnote{We note that the measured duration $t_{90}$ and accordingly the radiated energy $E_{\rm rad}$ may be underestimated in some events because of incomplete light curve coverage (some events were discovered only at the the end of the plateau phase and others were not observed until the end of the plateau; see Fig.~\ref{fig:obs}).}
Observed quantities are shown by circles or dot-dashed lines (when no information on $M_{\star}$ is available).  For comparison in Fig.~\ref{fig:L90t90} we show our light curve predictions (Fig.~\ref{fig:dist1}) assuming $M_{\rm ej} = \eta M_{\star}$ and $\bar{v}_{\rm E} = v_{\rm esc,\star} = (2GM_{\star}/R_{\star})^{1/2}$ for stars on the TAMS (red curves) and with radii 10 times their TAMS radius (10TAMS; blue curves), for a range of values $\eta = 10^{-2}$ (thinnest lines) to $\eta = 1$ (thickest lines).  For fixed $\eta$, the theoretical relations roughly follow the trends $L_{90}\propto M_{\star}$ and $t_{90}\propto M_{\star}^{1/2}$ predicted by Eqs.~(\ref{eq:tpl},\ref{eq:Lpl}).

Figure~\ref{fig:L90t90} shows that, in broad brush terms, our calculations can reproduce the observed luminosities $\sim 10^{38}-10^{40}$ erg s$^{-1}$ of most LRNe, and also the general trend of higher luminosity and longer duration for more massive stellar progenitors (e.g., \citealt{Kochanek14,Blagorodnova+21}).  However, there are several exceptions, including OT2011 in NGC4490 ($L_{90}\sim2\times10^{41}\,\rm erg\,s^{-1}$; \citealt{Pastorello+19a}), AT2014ej ($L_{90}\sim2\times10^{41}\,\rm erg\,s^{-1}$; \citealt{Stritzinger+20}), and AT2017jfs ($L_{90} \sim 3\times 10^{41}$ erg s$^{-1}$; \citealt{Pastorello+19b}), with luminosities that approach or exceed even the $\eta = 1$ TAMS prediction for very massive stars.  We return to this tension regarding the implied ejecta masses of some LRNe below.

We can determine the ejecta properties for individual LRNe by inverting our analytic formulae for the transient luminosity and duration (Eqs.~\ref{eq:tpl}, \ref{eq:Lpl}),
\begin{align}
M_{\rm ej}&\simeq 1.6\,\Msun\,f_{\rm ad,0.3}^{-1}\biggl(\frac{t_{\rm pl}}{100{\,\rm d}}\biggl)\biggl(\frac{L_{\rm pl}}{10^{39}\,\rm erg\,s^{-1}}\biggl)\ ,\label{eq:Mej_recombination}\\
\bar{v}_{\rm E}&\simeq480{\,\rm km\,s^{-1}}\rho_{\rm i,-11}^{-1/3}f_{\rm ad,0.3}^{-1/3}\biggl(\frac{t_{\rm pl}}{100\,\rm d}\biggl)^{-2/3}\biggl(\frac{L_{\rm pl}}{10^{39}{\,\rm erg\,s^{-1}}}\biggl)^{1/3}\ ,
    \label{eq:v_recombination1}
\end{align}
where we note that our calculations reveal that $f_{\rm ad,0.3}$ increases moderately $\sim 0.5-2$ with increasing stellar mass and radius (Fig.~\ref{fig:dist1}).  Eq.~\eqref{eq:v_recombination1} provides an independent velocity estimate from that obtained from spectroscopy of the transient, $v_{\rm obs}$, the consistency between which provides a check on the model (see below).

The ejecta launching radius $R_0$ can likewise be inferred from the observables in two ways:\\ (1) assuming $\bar{v}_{\rm E} = v_{\rm esc} = (2GM_{\star}/R_0)^{1/2}$ using the measured $M_{\star}$ when available, 
\begin{eqnarray}
R_0&\simeq& 5.4\,\Rsun\,f_{\rm ad,0.3}^{2/3}\rho_{\rm i,-11}^{2/3}\times \nonumber \\
&& \biggl(\frac{t_{\rm pl}}{100{\,\rm d}}\biggl)^{4/3}\biggl(\frac{L_{\rm pl}}{10^{39}\,\rm erg\,s^{-1}}\biggl)^{-2/3}\biggl(\frac{M_\star}{3\,\Msun}\biggl)\ ;\label{eq:R0_recombination}
\end{eqnarray}
(2) assuming $v_{\rm obs}=v_{\rm esc}$,
\begin{align}
R_0&\simeq13\,\Rsun\,\biggl(\frac{v_{\rm obs}}{300{\,\rm km\,s^{-1}}}\biggl)^{-2}\biggl(\frac{M_{\star}}{3\,\Msun}\biggl)\ 
\label{eq:R0_observed}
.
\end{align}

\begin{figure*}
\begin{center}
\includegraphics[width=160mm, angle=0,bb=0 0 501 486]{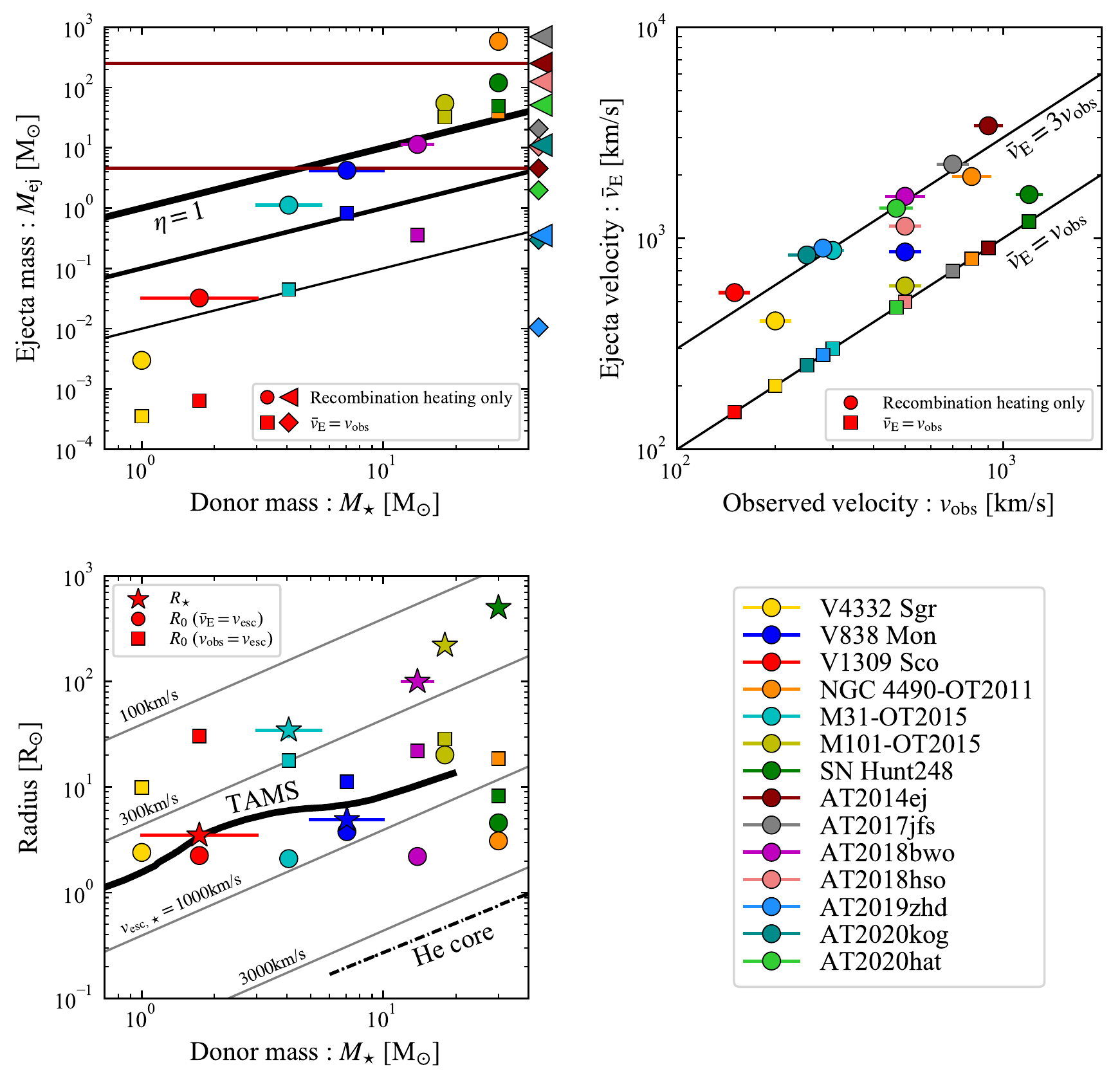}
\caption{For the same sample of LRNe from Fig.~\ref{fig:L90t90}, here we show: (top left panel) Ejecta mass $M_{\rm ej}$ estimated from light curve observables $\{L_{90}$, $t_{90}\}$ versus donor mass from pre-transient imaging, $M_{\star}$.  Circles show $M_{\rm ej}$ calculated from our light curve model assuming hydrogen recombination as the only energy source (Eq.~\ref{eq:Mej_recombination}), while squares show $M_{\rm ej}$ calculated from Eq.~\eqref{eq:tpl} assuming the mean ejecta speed equals that observed spectroscopically, $\bar{v}_{\rm E}=v_{\rm obs}$; triangles and diamonds show the same as circles and squares but for events without donor-mass constraint (horizontal lines denote the corresponding mass estimates for an event); for comparison, lines of constant $\eta = M_{\rm ej}/M_{\star} = 0.01,0.1, 1$ are shown as solid black lines; (top right) Ejecta velocity estimated from light curve observables (Eq.~\ref{eq:v_recombination1}) versus spectroscopic velocity, $v_{\rm obs}$ (Circles). Squares show the case when the estimated velocity is the same exactly as $v_{\rm obs}$.; (bottom left) Various radii versus donor mass.  Circles show the mass ejection radius $R_0$ estimated by equating the mean velocity $\bar{v}_{\rm E}$ inferred light curve modeling to the local escape speed
$v_{\rm esc} = (2GM_{\star}/R_0)^{1/2}$ (Eq.~\ref{eq:R0_recombination}), while squares show $R_0$ calculated assuming $v_{\rm esc}=v_{\rm obs}$ (Eq.~\ref{eq:R0_observed}).  Stars show the radius of the donor progenitor star, $R_{\star}$, inferred from pre-transient imaging.}
\label{fig:ejectavsstellarproperties}
\end{center}
\end{figure*}

Figure \ref{fig:ejectavsstellarproperties} compare the inferred ejecta properties $\{M_{\rm ej},\bar{v}_{\rm E},R_0\}$ to the stellar properties or transient properties $\{M_{\star},R_{\star},v_{\rm obs}\}$ for events for which the latter information is available (see also Table~\ref{table data}).  At a basic level most of our results for the baseline model (shown as circles or triangles in Fig.~\ref{fig:ejectavsstellarproperties}) make physical sense.  Inferred ejecta masses range from $\lesssim 10^{-2}\,\Msun$ in V4332 Sgr to $\sim 10\,\Msun$ in AT2018bwo, for example.  Likewise, the implied ejecta launching radii are all safely within the donor's interior, $R_{0} < R_{\star}$.  We return to what inferences can be drawn about the LRN population based on these data in Section \ref{sec:LRNpopulation}.

Upon closer inspection, several inconsistencies or tensions are also noticeable in Fig.~\ref{fig:ejectavsstellarproperties}.  Firstly, the implied ejecta masses from the most massive binary systems (e.g., NGC4490-OT2011 and AT2017jfs) are enormous $\gtrsim 100\,\Msun$ and greatly violate the theoretical limit $M_{\rm ej} > M_{\star} (\eta > 1)$ in which the entire mass of the donor is ejected, while a comparable number of events hover around $\eta \sim 1$ (see also \citealt{Soker20}).  A second point of tension is that the inferred ejecta velocities (estimated from Eq.~\ref{eq:v_recombination1}) systematically exceed those observed, $\bar{v}_{\rm E}\sim (2-3)\,v_{\rm obs}$.  Equivalently, the ejection radius $R_0$ one infers by equating $\bar{v}_{\rm E}$ to the local escape velocity (Eq.~\ref{eq:R0_recombination}) is smaller than that implied if the observed spectroscopic expansion velocity equals the escape speed at the ejection radius (squares in bottom panel of Fig.~\ref{fig:ejectavsstellarproperties}).

Keeping in mind that our formulae (Eqs.~\ref{eq:Mej_recombination} and \ref{eq:v_recombination1}) exhibit errors of a factor of a few, the latter mentioned velocity/radius discrepancy could in principle have several origins: (1) the observed spectroscopic velocities underestimate the true energy-weighted ejecta speed, for instance due how excitation and optical-depth affect what regions of the ejecta given lines probe; (2) inaccuracies in our light curve model cause an over-estimate of the plateau duration, which results in a large velocity for a given observed duration (see Eq.~\ref{eq:tpl});this could result from one or more of our simplifying assumptions, e.g.~spherical symmetry; (3) our model is missing physics, such as additional sources of the ejecta heating beyond that from hydrogen recombination. The latter possibility would also have implications for the inferred ejecta mass.

To illustrate this with an example, consider a hypothetical scenario in which the ejecta is heated during the plateau phase at a specific rate $\sim 10$ times larger than from hydrogen recombination (e.g., due to shock interaction with circumbinary material; Section \ref{sec:shocks}).  In such a case the true ejecta mass would be $\sim 10$ times smaller than inferred from Eq.~\eqref{eq:Mej_recombination} from the observed transient duration and luminosity; hence, for a fixed observed plateau duration $t_{\rm pl}$ (Eq.~\ref{eq:tpl}) the implied ejecta speed $\bar{v}_{\rm E} \propto M_{\rm ej}^{1/3}$ would be $\sim 10^{1/3} \sim 2$ times smaller than previously estimated neglecting the additional heating source, bringing it into closer agreement with $v_{\rm obs}$ (and likewise, the implied $R_0$ obtained assuming $\bar{v}_{\rm E} = v_{\rm esc}$ would be $\sim 4$ times smaller).  

To explore this possibility further, the top left panel of Fig.~\ref{fig:ejectavsstellarproperties} includes a second estimate of the ejecta mass (shown as squares and diamonds), which makes use of our expression for the plateau duration (Eq.~\ref{eq:tpl}) but supplants Eq.~\eqref{eq:Mej_recombination} (which assumes recombination as the sole energy source) with the assumption that $\bar{v}_{\rm E} = v_{\rm obs}$, i.e.~that the ejecta speed equals that observed from the transient spectroscopy.  Comparing the circles and squares (or triangles and diamonds), we observe that the required value of $M_{\rm ej}$ decreases significantly, bringing even the most luminous LRNe near or into the physically allowed region ($\eta \lesssim 1$).   Under the same assumption ($\bar{v}_{\rm E} = v_{\rm obs}$), the implied mass ejection radius increases but remains below the progenitor star surface for most events (squares in the bottom panel of Fig.~\ref{fig:obs}). 

To summarize, the ejecta masses one infers from the optical light curve assuming hydrogen recombination as the only energy source (as in our model) represent only upper limits.  If other sources of heating not accounted in our model are present, then (1) $M_{\rm ej}$ will be lower than this upper limit and; (2) the ejecta velocities $\bar{v}_{\rm E}$ implied by the observed LRN duration are brought into better agreement with those observed spectroscopically.

\begin{table*}
\begin{center}
\caption{LRNe Sample: Light curve observables, stellar progenitors, and inferred ejecta properties.}
\label{table data}
\begin{tabular}{lccccccccc}
\hline
Event&$L_{90}$&$t_{90}$&$E_{\rm rad}$&$v_{\rm obs}$$^\ddagger$&$M_\star$&$R_\star$&$M_{\rm ej}^\P$&$\bar{v}_{\rm E}^\P$&Ref.\\
&[erg/s]&[d]&[erg]&[km/s]&[$\Msun$]&[$\Rsun$]&[$\Msun$]&[km/s]&\\
\hline
V4332 Sgr&1.3e+37&15&1.7e+43&$200\pm20$&1&-&3.0e-03&410&1,2$^\dag$\\
V838 Mon&3.6e+39&79&2.4e+46&$500\pm50$&5-10&4-6&4.2e+00&860&3,4,5$^\dag$\\
V1309 Sco&8.9e+37&24&1.9e+44&$150\pm15$&1-3&3.5&3.2e-02&550&6,7$^\dag$\\
NGC 4490-OT2011&2.2e+41&180&3.3e+48&$800\pm100$&30&-&5.8e+02&2000&8,9$^\dag$\\
M31-OT2015&1.5e+39&50&6.4e+45&$300\pm20$&3-5.5&30-40&1.1e+00&880&10,11,12,13$^\dag$\\
M101-OT2015&1.4e+40&270&3.1e+47&$500\pm50$&18&220&5.5e+01&590&14\\
SN Hunt248&6.2e+40&130&6.9e+47&$1200\pm100$&30&500&1.2e+02&1600&9$^\dag$,15,16\\
AT 2014ej&2.2e+41&77&1.4e+48&$900\pm80$&-&-&2.5e+02&3400&17\\
AT 2017jfs&2.8e+41&170&4.0e+48&$700\pm70$&-&-&6.9e+02&2200&18\\
AT 2018bwo&1.3e+40&60&6.5e+46&$500\pm65$&12-16&100&1.1e+01&1600&19\\
AT 2018hso&4.5e+40&180&7.2e+47&$500\pm50$&-&-&1.2e+02&1100&20\\
AT 2019zhd&7.1e+38&33&2.0e+45&$280\pm30$&-&-&3.5e-01&900&21\\
AT 2020kog&6.6e+39&110&6.4e+46&$250\pm30$&-&-&1.1e+01&830&22\\
AT 2020hat&3.0e+40&110&2.9e+47&$470\pm50$&-&-&5.1e+01&1400&22\\
\hline
\multicolumn{10}{l}{
$^\ddagger$ Observed velocities are taken from Table 5 of \citet{Blagorodnova+21} for all events.}\\
\multicolumn{10}{l}{
$^\P$Obtained by Eqs.~\eqref{eq:Mej_recombination} and \eqref{eq:v_recombination1} for $\rho_{\rm i,-11}=f_{\rm ad,0.3}=1$.}\\
\multicolumn{10}{l}{
$^\dag$ References for the light curve data.}\\
\multicolumn{10}{l}{
{\bf Ref.} 1: \citet{Martini+99}, 2: \citet{Tylenda+05},
3: \citet{Munari+02}, 4: \citet{Bond+03},}\\
\multicolumn{10}{l}{5: \citet{Tylenda05}, 6: \citet{Mason+10}, 7: \citet{Tylenda+11}, 8: \citet{Smith+16}, 9: \citet{Pastorello+19b}}\\
\multicolumn{10}{l}{
10: \citet{Kurtenkov+15}, 11: \citet{Lipunov+17}, 12: \citet{Williams+15}, 13: \citet{MacLeod+17}}\\
\multicolumn{10}{l}{
14: \citet{Blagorodnova+17}, 15: \citet{Kankare+15}, 16: \cite{Mauerhan+15}, 17 : \citet{Stritzinger+20}}\\
\multicolumn{10}{l}{
18 : \citet{Pastorello+19a}, 19: \citet{Blagorodnova+21}, 20: \citet{Cai+19}, 21: \citet{Pastorello+21}}\\
\multicolumn{10}{l}{
22: \citet{Pastorello+21b}}\\
\end{tabular}
\end{center}
\end{table*}

\subsection{Comparison to SNe II Analytic Formulae}
\label{sec:Popovcompare}

The facts that recombination energy dominates the luminosity and gas pressure can exceed radiation pressure during the plateau phase (Fig.~\ref{fig:fiducial_peakquantities}), both violate implicit assumptions of analytic SNe light curve models \citep{Popov93,Sukhbold+16}.  Because several past works use these formulae to infer the ejecta properties from observed LRNe (e.g., \citealt{Ivanova+13b,MacLeod+17,Blagorodnova+21}), or predict observables in a population analysis (e.g., \citealt{Howitt+20}), it is important to quantify the implied inaccuracies.  

\begin{figure*}
\begin{center}
\includegraphics[width=160mm, angle=0,bb=0 0 500 229]{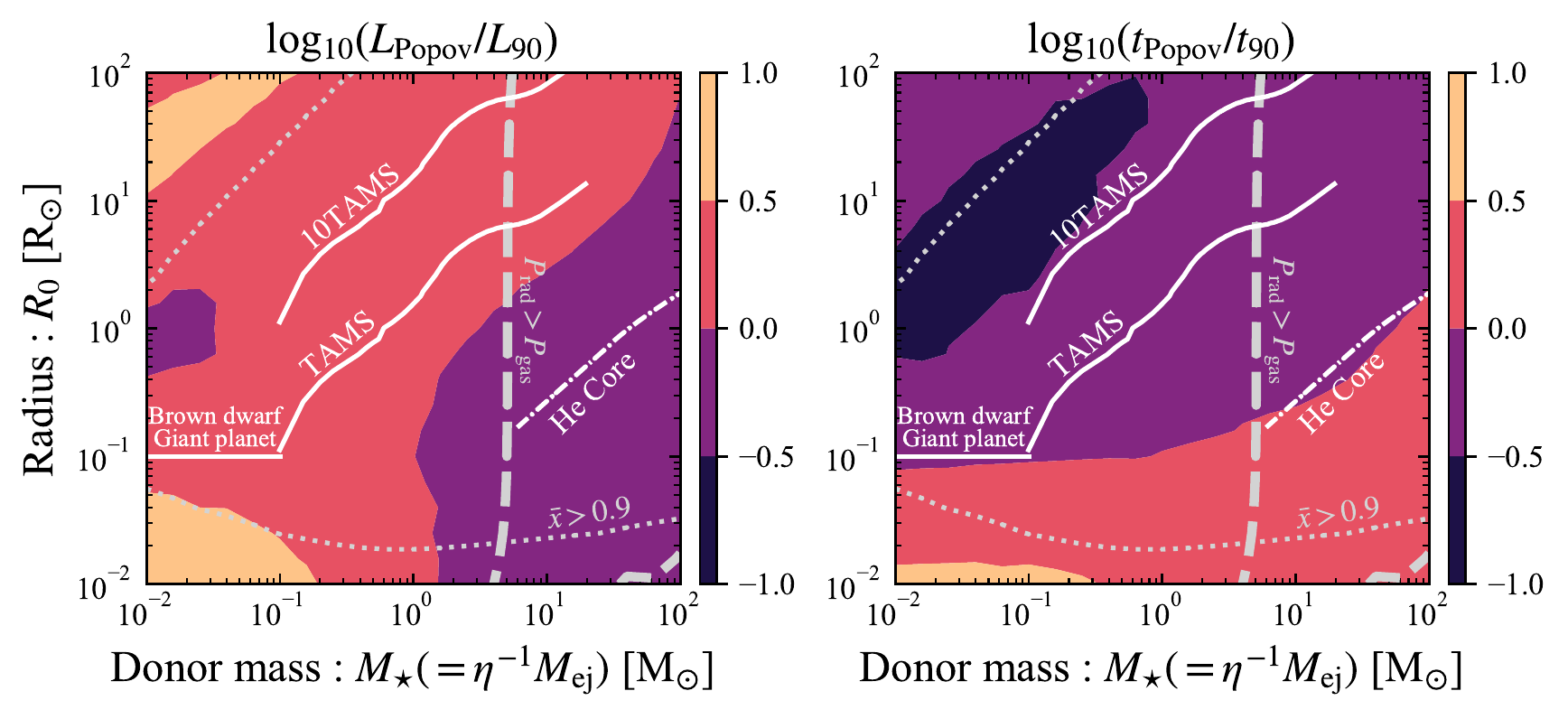}
\caption{The same as Fig.~\ref{fig:dist1} but showing the ratios of the transient luminosity and duration from our numerical calculations to the analytic formulae for Type II supernovae (Eqs.~\ref{eq:L_popov}, \ref{eq:t_popov}; \citealt{Popov93,Sukhbold+16}).  Along the gray dashed line contour along the radiation pressure is equal to the gas pressure at the time of ejection for the shell with $v=\bar{v}_{\rm E}$.  The assumption of radiation-dominated ejecta (e.g., \citealt{Popov93}) is violated to the left of this line.}
\label{fig:dist2}
\end{center}
\end{figure*}

Figure \ref{fig:dist2} shows the ratio of the transient luminosity and duration as extracted from our light curve calculation (Fig.~\ref{fig:dist1}) to those predicted by \citet{Popov93}.  For the latter we employ results calibrated using SN II radiative transfer calculations \citep{Sukhbold+16,Blagorodnova+21},
\begin{align}
L_{\rm Popov}&=1.3\times10^{39}{\,\rm erg\,s^{-1}\,}\times
    \nonumber\\
    &\biggl(\frac{M_{\rm ej}}{\Msun}\biggl)^{1/3}\biggl(\frac{R_0}{10\,\Rsun}\biggl)^{2/3}\biggl(\frac{\bar{v}_{\rm E}}{300\,\rm km\,s^{-1}}\biggl)^{5/3}\ ,
	\label{eq:L_popov}\\
t_{\rm Popov}&=51{\,\rm d\,}\biggl(\frac{M_{\rm ej}}{\Msun}\biggl)^{1/3}\biggl(\frac{R_0}{10\,\Rsun}\biggl)^{1/6}\biggl(\frac{\bar{v}_{\rm E}}{300\,\rm km\,s^{-1}}\biggl)^{-1/3}\ .
	\label{eq:t_popov}
\end{align}
Remarkably, Fig.~\ref{fig:dist2} shows that the calibrated Popov formulae (Eqs.~\ref{eq:L_popov},\ref{eq:t_popov}) agree with our model predictions across most of the relevant parameter space to within a factor $\lesssim 3$.  However, this agreement is largely a coincidence because the models are derived under completely different assumptions (Popov's formulae ``know'' nothing about hydrogen recombination energy, for instance).  The Popov formulae also exhibit different parameter dependencies from our model (Eqs.~\ref{eq:tpl},\ref{eq:Lpl}), however they (again, coincidentally) become similar for $\bar{v}_{\rm E} \sim v_{\rm esc}\propto(M_\star/R_0)^{1/2}$: $L_{\rm Popov}\propto M_{\rm ej}^{7/6}R_0^{-1/6}$ and $t_{\rm Popov}\propto M_{\rm ej}^{1/6}R_0^{1/3}$.  

As pointed out by \citet{MacLeod+17,Blagorodnova+21}, using the \citet{Popov93} expressions to infer ejecta properties from LRN observables typically result in large errors on the former.  To see why, consider inverting Eqs.~\eqref{eq:L_popov}, \eqref{eq:t_popov} to obtain the ejecta mass and launching radius as functions of plateau luminosity, duration, and ejecta velocity:
\begin{align}
M_{\rm ej}&\simeq18\,\Msun\,\biggl(\frac{t_{\rm pl}}{100\,\rm d}\biggl)^{4}\biggl(\frac{\bar{v}_{\rm E}}{300\,\rm km\,s^{-1}}\biggl)^{3}\biggl(\frac{L_{\rm pl}}{10^{39}\,\rm erg\,s^{-1}}\biggl)^{-1}\ ,
	\label{eq:MejPopov}\\
R_0&\simeq1.7\,\Rsun\,\biggl(\frac{t_{\rm pl}}{100\,\rm d}\biggl)^{-2}\biggl(\frac{\bar{v}_{\rm E}}{300\,\rm km\,s^{-1}}\biggl)^{-4}\biggl(\frac{L_{\rm pl}}{10^{39}\rm \,erg\,s^{-1}}\biggl)^{2}\ .
\label{eq:R0Popov}
\end{align}
The inferred values $M_{\rm ej}, R_0$ thus depend sensitively on the observables, particularly the ejecta velocity.  This is problematic because the measured velocity (e.g., as typically inferred from spectral line widths at particular epochs) may not coincide precisely with the bulk of the ejecta.  By contrast, the sensitivity of inferred ejecta quantities to LRN observables as predicted by our model (Eqs.~\ref{eq:Mej_recombination},\ref{eq:v_recombination1},\ref{eq:R0_recombination}), particularly if the donor star properties are independently constrained, allow for a more robust extraction of $\{M_{\rm ej},\bar{v}_{\rm E},R_0\}$, compared to Eqs.~(\ref{eq:MejPopov},\ref{eq:R0Popov}).  

\begin{figure*}
\begin{center}
\includegraphics[width=160mm, angle=0,bb=0 0 474 262]{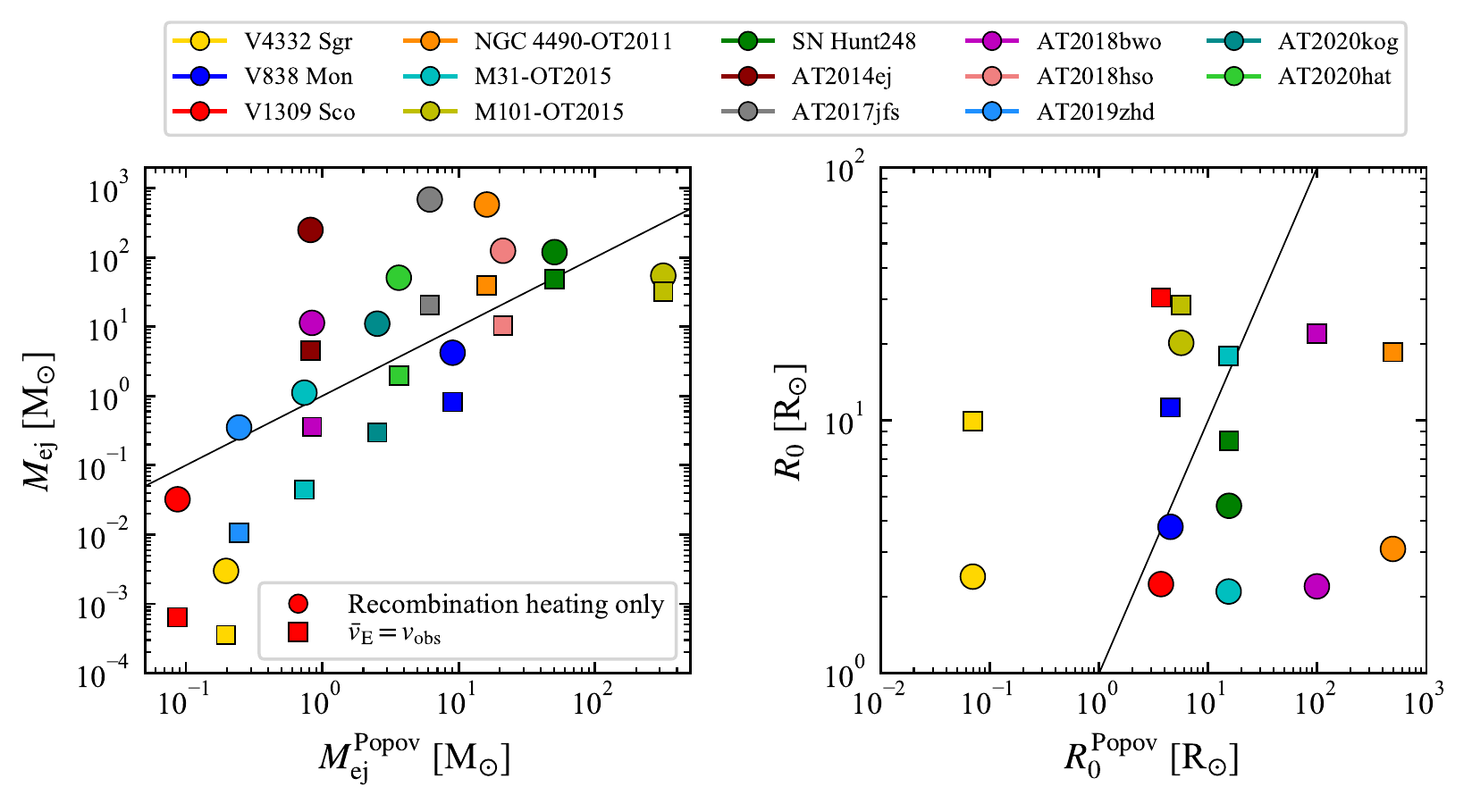}
\caption{For the same sample of LRNe in Fig.~\ref{fig:ejectavsstellarproperties}, here we compare the ejecta masses and launching radii inferred from our light curve model, $M_{\rm ej}/R_0$ on the vertical axis (using the two methods described in Fig.~\ref{fig:ejectavsstellarproperties} with the same symbols notation) to those obtained using the \citet{Popov93} supernova light curve formulae on the horizontal axis (Eqs.~\ref{eq:MejPopov}, \ref{eq:R0Popov}).
}
\label{fig:comparePopov}
\end{center}
\end{figure*}

Figure \ref{fig:comparePopov} compares the inferences on $\{M_{\rm ej},R_0 \}$ made employing our model (Table \ref{table data}) to those obtained by instead applying the modified \citet{Popov93} formulae (Eqs.~\ref{eq:MejPopov},\ref{eq:R0Popov}). We see that the ejecta masses implied by the Popov formulae can over- or under-estimate those of our model (circles) by more than 1-2 orders of magnitude.  In particular, the Popov formulae underestimate the implied $M_{\rm ej}$ for the most luminous LRNe (largest $M_{\rm ej}$ events), masking the $M_{\rm ej} > M_{\star}$ tension implied by our modeling (Sec.~\ref{sec:IndividualLRN}).
In applying the Popov formulae here, we have estimated the ejecta mass and radius using the observed spectroscopic  velocity $\bar{v}_{\rm E}=v_{\rm obs}$, which typically differs from the surface escape speed (Fig.~\ref{fig:progenitor}). The sensitive dependence of the Popov formulae velocity thus accounts for the larger deviations of the ejecta properties from our estimates in Fig.~\ref{fig:comparePopov} than naively suggested by Fig.~\ref{fig:dist2} which instead assumed $\bar{v}_{\rm E}=v_{\rm esc}$.

\section{Discussion}
\label{sec:discussion}

\subsection{Shock Interaction as a Luminosity Source}
\label{sec:shocks}

Hydrogen recombination typically dominates the radiated energy in our baseline model because the ejecta's initial thermal energy is mostly lost to adiabatic expansion before it can be radiated.  However, some of this thermal energy can be regenerated if the merger ejecta were to collide with slower material at larger radii $\gg R_{\star}$ surrounding the binary \citep{Metzger&Pejcha17}.  This pre-existing slower material could arise from an earlier episode of mass-transfer leading up to the dynamical merger event, for instance a brief thermal-timescale mass-transfer phase (e.g., \citealt{Blagorodnova+21}) or other physical processes generating mass-loss from the binary's $L_2$ point \citep{Pejcha+16a,Pejcha+16b,Pejcha+17}.  The presence of pre-dynamical ejecta is supported observationally by the slow early rise in the pre-maximum light curves of LRNe such as V1309 Sco \citep{Pejcha14,Pejcha+17} and the redder than expected early-time colors of V838 Mon and AT2018bwo \citep{Blagorodnova+21}. 

The characteristic mass $M_{\rm pre}$ of the slowly-expanding (velocity $v_{\rm pre}$) material surrounding the binary at the time of merger is uncertain, but in many cases is expected to be smaller than the ejecta mass released at later times, $M_{\rm ej}$ (e.g., at most $M_{\rm pre} \sim 0.1M_{\star}$ is predicted via $L_{2}$ mass-loss; \citealt{MacLeod+17,Pejcha+17,MacLeod&Loeb20}; however, see \citealt{Blagorodnova+21}).  Assuming $M_{\rm pre} \ll M_{\rm ej}$ and $v_{\rm pre} \ll \bar{v}_{\rm E}$, and that the shock-interaction is momentum conserving, the thermal energy released by the collision is given by
\begin{equation}
E_{\rm sh} \simeq \frac{1}{2}M_{\rm pre}\bar{v}_{\rm E}^{2}\ .
\end{equation}
This exceeds the energy released by hydrogen recombination, $\varepsilon_{\rm H}XM_{\rm ej}/\MP$, for 
\begin{equation}
\frac{M_{\rm pre}}{M_{\rm ej}} \gtrsim \frac{2X\varepsilon_{\rm H}}{\MP \bar{v}_{\rm E}^{2}} \sim 0.02\left(\frac{\bar{v}_{\rm E}}{300\,{\rm km\,s^{-1}}}\right)^{-2}
\end{equation}
Thus, if even a few percent of the ejecta mass is present on radial scales $R_{\rm pre} \sim \bar{v}_{\rm E} t_{\rm pl} \sim 10^{14}-10^{15}$ cm prior to the merger, the shock will compete with recombination in powering the light curve.  A related, alternative heating mechanism for the LRN ejecta is a fast outflow or jet powered by accretion onto the accretor star or donor star core (e.g., \citealt{Soker&Kashi16,Soker20,Soker&Kaplan21}).

If the duration of the transient is still set by hydrogen recombination in this scenario, then the transient duration is still given by $t_{90} \sim t_{\rm pl} \propto M_{\rm ej}^{1/3}\bar{v}_{\rm E}^{-1}$ (Eq.~\ref{eq:tpl}).  However, instead of Eq.~\eqref{eq:Epl} for the plateau luminosity, the shock scenario predicts $E_{\rm pl} \propto E_{\rm sh} \propto \bar{v}_{\rm E}^{2}$ and hence
\begin{align}
L_{\rm pl}^{\rm sh} &\sim \frac{E_{\rm sh}}{t_{\rm pl}} \simeq 7\times 10^{39}{\rm erg\,s^{-1}}\rho_{\rm i,-11}^{1/3}\times
    \nonumber\\
&\left(\frac{M_{\rm pre}}{0.1M_{\rm ej}}\right)\left(\frac{\bar{v}_{\rm E}}{300\,{\rm km\,s^{-1}}}\right)^{3}\left(\frac{M_{\rm ej}}{\Msun}\right)^{2/3}.
\label{eq:Lsh}
\end{align}
The value of $L_{\rm pl}^{\rm sh}$ can now reach $\gtrsim 10^{41}$ erg s$^{-1}$ for events like AT2017jfs which possess a relatively high velocity $v_{\rm obs} \sim 700$ km s$^{-1},$ and which otherwise required unphysically large ejecta masses when considering recombination energy alone (Fig.~\ref{fig:ejectavsstellarproperties}).  Depending on its opacity $\kappa$, the optical depth through the circumbinary material,
\begin{eqnarray}
\tau_{\rm pre} &\sim& \frac{\kappa M_{\rm pre}}{4\pi R_{\rm pre}^{2}} \nonumber \\
&\sim& 1.6\left(\frac{\kappa}{0.1\,{\rm cm^{2}\,g^{-1}}}\right)\left(\frac{M_{\rm pre}}{0.1\,\Msun}\right)\left(\frac{R_{\rm pre}}{10^{15}\,{\rm cm}}\right)^{-2} \ ,
\end{eqnarray}
is sufficiently large to redden the early transient emission (e.g., \citealt{Blagorodnova+21}).  Future work is required to incorporate the time-dependent shock luminosity, given an assumed density profile of the pre-merger cirumbinary medium (e.g., \citealt{Metzger&Pejcha17}), self-consistently into our light curve model framework (Section \ref{sec:method}).

One might also expect shock interaction to manifest through observational signatures other than just the light curve, such as optical emission lines and thermal X-rays from hot shocked gas and non-thermal radio synchrotron emission from relativistic electrons accelerated at the shock (as seen from internal shocks in classical novae, for example; \citealt{Chomiuk+21}).  However, the equatorial geometry of the slow circumbinary medium (e.g., \citealt{Pejcha+16a}) implies that the polar regions of the merger ejecta and hence the optical photosphere will quickly ``wrap around'' the equatorial shock, blocking signatures of the latter from most viewing angles (see \citealt{Andrews&Smith18} for an example of this in the context of shock-powered Type II supernovae).  At late times after recombination completes, the opacity will drop, the photosphere recedes and residual shock signatures may become visible; however, the high optical-wavelength opacity once dust formation begins around the same time may keep these inner regions opaque and continue to obscure shock-related signatures.

\subsection{Inferences about the LRN Population}
\label{sec:LRNpopulation}

A goal of measuring the ejecta properties for a sample of LRNe is to deduce information about the physics of the merger process as well as the demographics of merging binary systems. 

Putting aside the ambiguities that arise in measuring individual ejecta masses in the face of additional sources of ejecta heating, the top panel of Fig.~\ref{fig:ejectavsstellarproperties} reveals an interesting trend that is likely to remain robust in spite of these uncertainties: the fraction of the stellar envelope being ejected $\eta = M_{\rm ej}/M_{\star}$ systematically increases with more massive donor stars.  This arises because the average luminosity of LRNe increases sharply with stellar mass, $L \propto M_{\star}^{\alpha}$ with $\alpha \sim 2-3$ (e.g., \citealt{Kochanek14}), while recombination powered models predict a sub-linear dependence of $L_{\rm pl} \propto M_{\rm ej}^{2/3}$ (Eq.~\ref{eq:Lpl}).  What could give rise to the increasing $\eta(M_{\star})$ trend?  

The fraction of the donor's star envelope unbound during the merger process is likely to increase with secondary mass and hence the mass ratio of the binary (e.g., \citealt{Nordhaus+10}).  The fact that the average mass ratio of interacting binaries is expected to increase with stellar mass \citep{Moe&DiStefano17} could therefore contribute to $\eta$ increasing with $M_{\star}$. Another contributing factor is that more massive LRN progenitors appear to be further evolved off the main sequence (Fig.~\ref{fig:progenitor}): more evolved stars possess a more pronounced core-envelope structure, facilitating the ejection of a higher fraction of the envelope as the accretor releases more gravitational energy spiraling into the compact core. A smaller ejection radius also increases the ejecta velocity, which acts to increase the transient luminosity (Eq.~\ref{eq:Lpl}).
Indeed, the most luminous events tend to have larger observed velocities (see Table~\ref{table data}).

The current LRN sample could also be biased by a selection effect: since the intrinsic nuclear luminosity of the star at all phases of evolution increases steeply with stellar mass (dotted and dashed lines in Fig.~\ref{fig:L90t90}), low-$M_{\rm ej}$ (low-$\eta$) events could be selected against in their discovery because of the lower contrast between the transient and star luminosity. Moreover, since massive progenitors are rare, their outbursts are more typically detected as extragalactic transients; hence, it is natural to miss low-luminosity events from massive stars. These would imply the existence of a hidden population of low-$M_{\rm ej}$ mergers (``minor mergers'') from massive stars. This hidden population would need to be taken into account when comparing the $M_{\star}$-dependent rate of LRN to the initial mass function of stars or to rate predictions from binary population synthesis modeling (e.g., \citealt{Kochanek14,Howitt+20}). 

\subsection{Stellar Mergers with Planets and Compact Objects}
\label{sec:othermergers}

Our model can be applied to predict the optical light curves from other types of mergers involving mass ejection from hydrogen-rich bodies.

One example is the merger of a giant planet or brown dwarf with a more massive host companion star (e.g., \citealt{Metzger+12}).  When the average density of the planet is larger than that of the star, the energy released as the planet plunges slowly below the stellar surface can give rise to mass ejection (and a longer-term evolution of the star's luminosity due to energy released deeper in the stellar envelope; \citealt{Metzger+17,MacLeod+18b}).  Alternatively, if the planet is less dense than the star (e.g., ``inflated'' Hot Jupiters; \citealt{Baraffe+10}), then it can be tidally disrupted above the surface, forming an accretion disk around the star.  Outflows from the resulting super-Eddington accretion flow are likely to unbind a significant fraction of its mass from the planet (e.g., \citealt{Metzger+12}).  In both cases then, a fraction of the planet/brown dwarf mass $\sim 10^{-3}-10^{-2}\,\Msun$ is ejected from radii $R_0 \sim R_{\star}$ at close to the surface escape speed.  

Figure~\ref{fig:L90t90} shows that for $M_{\star} \sim\Msun$ and $\eta \sim 10^{-3}-10^{-2}$, the predicted transient lasts days to weeks with a luminosity $L \sim 10^{37}-10^{38}$ erg s$^{-1}$.  These light curve properties are comparable to those of classical novae (e.g., \citealt{Strope+10}), with which planet-star mergers could easily be confused.  Based on the decay rate of observed gas giant planets orbits due to tidal dissipation, \citet{Metzger+12} estimate a planet-star merger rate of $\sim 0.1-1$ per year in the Milky Way, consistent (within an order of magnitude) of the stellar merger rate \citep{Kochanek14} and $\sim 100$ times lower than the rate of classical novae (e.g., \citealt{Kawash+21}).  Given their low luminosities, and low relative frequency to classical novae, it is unsurprising that no confirmed planet-star mergers have yet been discovered.

Our model can also be applied to mergers of hydrogen-rich stars with compact objects, such as white dwarfs (WD; e.g., \citealt{Shara&Shaviv77,Michaely&Shara21,Metzger+21}), neutron stars (NS), or stellar- or intermediate-mass black holes (BH; e.g., \citealt{Perets+16,Fragione+20,Kremer+21,Kremer+22}), which give rise to up to several solar masses of ejecta expanding at velocities of hundreds to thousands of km s$^{-1}$.  Due to the deeper potential well of the WD/NS/BH, the kinetic energies of the ejecta from these events can exceed those of ordinary star mergers; however, the small ejection radii$-$less than the tidal disruption radius, itself typically comparable to the stellar radius$-$imply that most of the initial thermal energy will be lost to adiabatic expansion, as in the stellar mergers we have studied.  As a result, the total radiated optical energy from the unbound ejecta will also be limited to that from its hydrogen recombination, i.e.~$\lesssim f_{\rm ad}\varepsilon_{\rm H}X\Msun/\MP \sim 10^{46}$ erg per solar mass of ejecta.  The predicted luminosity from star-WD/NS/BH mergers are thus unlikely to exceed $\sim 10^{39}-10^{40}$ erg s$^{-1}$ \citep{Metzger+21}, less than estimates that do not account for adiabatic losses prior to recombination (e.g., \citealt{Michaely&Shara21}).  Of course, if as may be the case also in stellar mergers (Sec.~\ref{sec:shocks}), star-WD/NS/BH mergers are accompanied by an extra central luminosity source which re-energizes the unbound ejecta (e.g., internal shocks due to ongoing outflows from the central engine; e.g., \citealt{Dexter&Kasen13,Margalit&Metzger16,Gottlieb+22}) then the transient luminosities could be considerably higher than from recombination power alone.

\section{Conclusions}
\label{sec:conclusions}

We have presented a new model for the light curves of LRNe from stellar mergers, which self-consistently accounts for the range of relevant opacities, an arbitrary degree of radiation and gas pressure, and the energy released from hydrogen recombination.  We apply our model to a sample of 14 LRNe (Fig.~\ref{fig:obs}, Table \ref{table data}) to infer the ejecta properties (mass, velocity, and launching radius) and compare to the mass/radios of the progenitor donor star when available (Fig.~\ref{fig:progenitor}).  Our conclusions can be summarized as follows:
\begin{itemize}

    \item Our model naturally reproduces the observation (Fig.~\ref{fig:obs}) that 
    LRN light curves typically possess two luminosity peaks: (1) a short initial peak typically lasting days powered by the initial thermal energy (``cooling emission'') from the outermost fastest layers of the ejecta; and (2) a longer peak or plateau lasting weeks or months, powered by energy released from hydrogen recombination from the bulk of the mass.  While radiation pressure can dominate in the ejecta during the first peak, gas pressure typically dominates the latter plateau, at odds with the common assumption of supernova light curve models (e.g., \citealt{Popov93}).
    
    The potential for two light curve peaks was recognized previously as a result of distinct physical stages during the merger \citep{MacLeod+17,Metzger&Pejcha17}.  Here, we emphasize (again) that such a double-peaked feature results due to the stratification of even a single spherically symmetric ejecta shell into layers with distinct velocities and initial thermal energy content.  Multiple distinct mass-loss episodes (e.g., \citealt{Ivanova+13b,Pastorello+19a}) are not required to generate multiple light curve peaks.   
    
    \item The luminosity and duration of the early and late light curve peaks follow from simple analytic estimates (Eqs.~\ref{eq:tpk},\ref{eq:Lpk} and Eqs.~\ref{eq:tpl},\ref{eq:Lpl}, respectively).  These formulae reproduce within a factor of a few the average luminosity $L_{90}$ and duration $t_{90}$ (over which 90\% of the total optical energy is received) across a wide parameter space, corresponding to mergers spanning the main sequence and post-main sequence phases of stars of mass $\sim 0.1\,\Msun$ to $\sim 100\,\Msun$ (Figs.~\ref{fig:ratio_ourmodel},\ref{fig:L90t90}).  A one-zone calculation, similar in methodology to \citet{Popov93} but assuming gas pressure-dominated conditions and accounting for recombination energy, reproduces similar scalings (Appendix \ref{sec:appendix}).  
    
    The total radiated energy $E_{\rm rad}$ is dominated by the recombination-powered plateau; hence the measured value of $E_{\rm rad}$ is proportional to the ejecta mass (up to an adiabatic loss factor $f_{\rm ad} \sim 0.1$; Fig.~\ref{fig:dist1}).
    
    \item We determine the ejecta properties $\{M_{\rm ej}, \bar{v}_{\rm E}, R_0\}$ of individual LRN events, first under the assumption that hydrogen recombination is the sole source of ejecta heating after launch, and compare them to the masses and radii of observed stellar progenitors of LRNe (Figs.~\ref{fig:L90t90}, \ref{fig:ejectavsstellarproperties}, Table \ref{table data}).  The derived ejecta properties differ from those derived by applying SN II light curve models such as \citet{Popov93} (e.g., Figs.~\ref{fig:dist2}, \ref{fig:comparePopov}). The latter also come with larger uncertainties due to the greater sensitivity of the formulae to the observables (Eqs.~\ref{eq:t_popov},\ref{eq:L_popov}).
    Remarkably, our results roughly agree with those by \citet{Popov93} under the assumption that the ejecta velocity equals the stellar escape velocity (Figs.~\ref{fig:dist2}). However, such information is not typically available in the Popov framework, and instead using the observed spectroscopic velocity results in drastically different ejecta property estimates compared to those obtained applying our model (Fig.~\ref{fig:comparePopov}).
        
    \item In broad terms, for the range of expected ejecta masses and velocities of stars at different stages of stellar evolution, we predict merger light curve durations $\sim$ weeks to months and luminosities $\sim 10^{38}-10^{40}$ erg s$^{-1}$, consistent with observed LRNe, as well as a general trend of increasing luminosity with increasing stellar mass (Fig.~\ref{fig:L90t90}).  
    
    \item However, several LRNe arising from mergers involving the most massive stars $\gtrsim 10\,\Msun$, exceed the luminosity upper limit corresponding to ejection of the entire donor star envelope ($M_{\rm ej}/M_{\star} = \eta > 1$; circles in Figs.~\ref{fig:L90t90} and \ref{fig:ejectavsstellarproperties}).  The inferred mean ejecta speed $\bar{v}_{\rm E}$ of many events also exceed those inferred by transient spectroscopy by a factor of $\sim 3$.  
    
    Both the luminosity and velocity tension can be alleviated by postulating the existence of an additional energy source for the ejecta exceeding that from hydrogen recombination (squares in Fig.~\ref{fig:ejectavsstellarproperties}).  This reduces the required ejecta mass to explain the radiated energy as well as the ejecta velocity needed to explain the light curve duration $t_{\rm pl} \propto M_{\rm ej}^{1/3}\bar{v}_{\rm E}^{-1}.$
    
    \item
    A promising source for the required ``extra'' heating is shock-interaction with circumbinary material generated prior to the dynamical phase of the merger (e.g., \citealt{Metzger&Pejcha17}).  The strong velocity dependence of shock-generated heating $\propto v^{2}$ and its associated plateau luminosity (Eq.~\ref{eq:Lsh}), coupled with the observation that the ejecta velocities of LRNe increase with stellar mass (Fig.~\ref{fig:progenitor}), would together act to steepen the $L-M_{\star}$ relationship relative to the fiducial recombination-only prediction (Fig.~\ref{fig:L90t90}), bringing it closer into alignment with LRN observations which find $L \propto M_{\star}^{2-3}$ (\citealt{Kochanek14}; Fig.~\ref{fig:L90t90}).
    
LRN donor star progenitors with larger masses appear systematically more evolved off the main sequence and generate merger ejecta with velocities exceeding the surface escape speed (Fig.~\ref{fig:progenitor}); this could arise due to the better developed core-envelope structure of evolved stars, which enables the secondary to inspiral deeper into the gravitational potential well.  

\item Although the robust conclusions we are able to draw are limited by the modest statistics afforded by the current sample of stellar merger transients, this population is expected to grow considerably in the years ahead.  \citet{Howitt+20} predict that the {\it Vera Rubin Observatory} will discover $\sim 500$ LRNe per year in nearby galaxies, for which pre-transient imaging will enable progenitor star constraints, allowing the methodology presented here to be greatly expanded.   
    
    \item Our model can also be generalized to other merger events involving the ejection of hydrogen-rich material, such as planet-star mergers and star-compact object mergers.  However, the relatively modest luminosities achieved by hydrogen recombination power alone, coupled with the comparatively low rates of these events, will likely render these transient signals challenging to detect.  
    
\end{itemize}

\acknowledgements
We are in debt to Nadia Blagorodnova for providing useful guidance on assembling LRN light curve data.  We thank Ondrej Pejcha for helpful conversations and comments on an early draft of the text.  This work is supported in part by JSPS Overseas Research Fellowships (T.M.).  B.D.M. acknowledges support from the National Science Foundation (grant number AST-2009255).

\appendix

\section{Analytical light curve model}
\label{sec:appendix}

We present a one-zone model \citep{Dexter&Kasen13,Matsumoto+16} for recombination powered transients analogous to \citet{Popov93} but for which the ejecta pressure is dominated by gas instead of radiation and accounting for the energy released by recombination.
We again consider a spherical and homologously expanding ejecta.
The initial ejecta temperature is roughly given by equating the kinetic energy $M_{\rm ej}v^2/2$ with internal energy $3M_{\rm ej}\KB T/2\MP$:
\begin{align}
T_0\simeq\frac{\MP v^2}{3\KB}\simeq3.6\times10^{6}{\,\rm K\,}\biggl(\frac{v}{300\,\rm km\,s^{-1}}\biggl)^2\ .
\end{align}
The ejecta is dominated by gas pressure for initial densities larger than a critical value,
\begin{align}
\rho_0\simeq 1.4{\,\rm g\,cm^{-3}\,}\biggl(\frac{M_{\rm ej}}{\Msun}\biggl)\biggl(\frac{R_0}{\Rsun}\biggl)^{-3}\gtrsim1.5\times10^{-3}{\,\rm g\,cm^{-3}\,}\biggl(\frac{v}{300{\,\rm km\,s^{-1}}}\biggl)^{6}\ .
\end{align}
The time evolution of temperature is determined by the first law of thermodynamics:
\begin{align}
\frac{dE}{dt}=-P\frac{dV}{dt}-L\ ,
	\label{eq thermodynamics}
\end{align}
where $E$, $P$, $V$, and $L$ are the internal energy, pressure, volume, and luminosity, respectively.
Since the gas-pressure dominates the radiation pressure, the internal energy and pressure are given by $E\sim 3M_{\rm ej}\KB T/2\MP$ and $P=(\gamma_3-1)E/V$ with the adiabatic index $\gamma_3=5/3$ as long as the recombination is negligible.
The radiative luminosity is given by the diffusion approximation:
\begin{align}
L\simeq 4\pi R^2\frac{c}{3\kappa \rho}\frac{d(aT^4)}{dR}\sim 4\pi R^2\frac{acT^4}{3\kappa \rho R}\ ,
	\label{eq diffusion approximation}
\end{align}
where $\rho=M_{\rm ej}/V$ is the density.
Combining Eq. \eqref{eq thermodynamics} with \eqref{eq diffusion approximation}, we have 
\begin{align}
\frac{dT}{dR}&=-\frac{2T}{R}-{\cal A}R^4T^4\ ,
\end{align}
where ${\cal A}={32\pi^2ac\MP}/({27\KB \kappa M_{\rm ej}^2}v)$.
Assuming the opacity is constant, we integrate this equation to obtain
\begin{align}
\frac{T}{T_0}&=\biggl[1+\xi_0\biggl(1-\frac{R_0}{R}\biggl)\biggl]^{-1/3}\biggl(\frac{R}{R_0}\biggl)^{-2}\ ,
    \label{eq:append T}\\
\xi_0&=3{\cal A}R_0^5T_0^3\simeq2.0\times10^{-10}\,\kappa_{0.32}^{-1}\biggl(\frac{M_{\rm ej}}{\Msun}\biggl)^{-2}\biggl(\frac{R_0}{\Rsun}\biggl)^5\biggl(\frac{v}{300{\,\rm km\,s^{-1}}}\biggl)^5\ ,
\end{align}
where $\kappa=0.32\,\kappa_{0.32}\,\rm cm^2\,g^{-1}$.
For a typical gas-pressure-dominated ejecta, $\xi_0\ll1$ and the temperature and luminosity decrease as $T\propto t^{-2}$ and $L\propto t^{-4}$, respectively.

When the temperature drops to a value $T_{\rm i} \sim 10^{4}$ K, recombination begins in the outermost layers of the ejecta.  As time proceeds, the recombination layer recedes back into the ejecta shell. We denote the location of the recombination front by a dimensionless radial coordinate $x\equiv R_{\rm i}/R$ and assume that the gas temperature is fixed to $T_{\rm i}$ at the layer.
Considering now only the volume inside the recombination front $4\pi (xR)^3/3$, the diffusion luminosity (Eq.~\ref{eq diffusion approximation}) becomes
\begin{align}
L\simeq 4\pi (xR)^2\frac{c}{3\kappa \rho}\frac{d(aT^4)}{dR}\sim 4\pi (xR)^2\frac{acT_{\rm i}^4}{3\kappa \rho xR}\ .
\end{align}
Then Eq.~\eqref{eq thermodynamics} gives the time evolution of the recombination front $x$:
\begin{align}
\frac{dx}{dR}=-\frac{\lambda x}{R}-\frac{(1-\lambda){\cal A}T_{\rm i}^3R^4}{3x}\ ,
\end{align}
where we define $\lambda\equiv (\gamma_3-1)/\gamma_3$.
Assuming $\lambda$ and $\kappa$ are constant during the recombination phase, we can integrate this equation analytically to obtain
\begin{align}
x&=\biggl[1-\frac{2(1-\lambda)\xi_{\rm i}}{9(2\lambda+5)}\biggl\{\biggl(\frac{R}{R_{\rm i}}\biggl)^{2\lambda+5}-1\biggl\}\biggl]^{1/2}\biggl(\frac{R}{R_{\rm i}}\biggl)^{-\lambda}\ ,\\
\xi_{\rm i}&=3{\cal A}R_{\rm i}^5T_{\rm i}^3\simeq1.0\times10^{-11}\,\kappa_{0.32}^{-1}T_{\rm i,4}^{1/2}\biggl(\frac{M_{\rm ej}}{\Msun}\biggl)^{-2}\biggl(\frac{R_0}{\Rsun}\biggl)^5\biggl(\frac{v}{300{\,\rm km\,s^{-1}}}\biggl)^4\ ,
\end{align}
where $T_{\rm i}=10^4\,T_{\rm i,4}\,\rm K$ and the ejecta radius at the beginning of the recombination is given by Eq.~\eqref{eq:append T},
\begin{align}
R_{\rm i}\simeq R_0\biggl(\frac{T_0}{T_{\rm i}}\biggl)^{1/2}\simeq 19\,R_0\,T_{\rm i,4}^{-1/2}\biggl(\frac{v}{300{\,\rm km\,s^{-1}}}\biggl)\ .
\end{align}
During the recombination phase, the light curve shows a plateau similar to that predicted by \citet{Popov93}, with its duration given by setting $x=0$:
\begin{align}
t_{\rm pl}=\frac{R_{\rm i}}{v}\biggl(\frac{9(2\lambda+5)}{2(1-\lambda)\xi_{\rm i}}\biggl)^{\frac{1}{2\lambda+5}}
\overset{\lambda=0}{\simeq}150{\,\rm d\,}\kappa_{0.32}^{1/5}T_{\rm i,4}^{-3/5}\biggl(\frac{M_{\rm ej}}{\Msun}\biggl)^{2/5}\biggl(\frac{v}{300{\,\rm km\,s^{-1}}}\biggl)^{-4/5}\ .
    \label{eq:appen tpl}
\end{align}
In the last equality, we have set $\lambda=0$ because the temperature does not evolve significantly during the plateau.  A characteristic luminosity may be given by the peak of the plateau, which is obtained by setting $dL/dt=0$,
\begin{align}
L_{\rm pl}=\frac{16\pi^2acT_{\rm i}^4R_{\rm i}^4}{9\kappa M_{\rm ej}}\biggl(\frac{2\lambda+5}{13}\biggl)^{1/2}\biggl(\frac{9(4-\lambda)(2\lambda+5)}{13(1-\lambda)\xi_{\rm i}}\biggl)^{\frac{4-\lambda}{2\lambda+5}}
\overset{\lambda=0}{\simeq}6.0\times10^{38}{\,\rm erg\,s^{-1}\,}\kappa_{0.32}^{-1/5}T_{\rm i,4}^{8/5}\biggl(\frac{M_{\rm ej}}{\Msun}\biggl)^{3/5}\biggl(\frac{v}{300\,\rm km\,s^{-1}}\biggl)^{4/5}\ .
    \label{eq:appen Lpl}
\end{align}
The total radiated energy is then given by
\begin{align}
E_{\rm pl}\sim L_{\rm pl}t_{\rm pl}
\overset{\lambda=0}{\simeq}4.4\,\frac{\KB T_{\rm i} M_{\rm ej}}{\MP}\simeq7.8\times10^{45}{\,\rm erg\,}T_{\rm i,4}\biggl(\frac{M_{\rm ej}}{\Msun}\biggl)\ .
    \label{eq:appen Epl}
\end{align}
For $\lambda=0$ ($\gamma_3 = 1$) and $T_{\rm i,4} \sim 1$ we see that the plateau energy matches that from hydrogen recombination (Eq.~\ref{eq:Epl}) and the parameter dependencies for $L_{\rm pl} \propto M_{\rm ej}^{3/5}v^{4/5}$ (Eq.~\ref{eq:appen Lpl}), $t_{\rm pl} \propto M_{\rm ej}^{2/5}v^{-4/5}$ (Eq.~\ref{eq:appen tpl}) and their normalizations, are similar to those we derived in Eqs.~\eqref{eq:tpl}-\eqref{eq:Lpl} under different assumptions in Sec.~\ref{sec:example}, namely $L_{\rm pl} \propto M_{\rm ej}^{2/3}v$ and $t_{\rm pl} \propto M_{\rm ej}^{1/3}v^{-1}$.

\bibliographystyle{aasjournal}
\bibliography{refs}

\begin{thebibliography}{}
\expandafter\ifx\csname natexlab\endcsname\relax\def\natexlab#1{#1}\fi
\providecommand{\url}[1]{\href{#1}{#1}}
\providecommand{\dodoi}[1]{doi:~\href{http://doi.org/#1}{\nolinkurl{#1}}}
\providecommand{\doeprint}[1]{\href{http://ascl.net/#1}{\nolinkurl{http://ascl.net/#1}}}
\providecommand{\doarXiv}[1]{\href{https://arxiv.org/abs/#1}{\nolinkurl{https://arxiv.org/abs/#1}}}

\bibitem[{{Andrews} \& {Smith}(2018)}]{Andrews&Smith18}
{Andrews}, J.~E., \& {Smith}, N. 2018, \mnras, 477, 74,
  \dodoi{10.1093/mnras/sty584}

\bibitem[{{Arnett}(1980)}]{Arnett80}
{Arnett}, W.~D. 1980, \apj, 237, 541, \dodoi{10.1086/157898}

\bibitem[{{Banerjee} {et~al.}(2015){Banerjee}, {Nuth}, {Misselt}, {Varricatt},
  {Sand}, {Ashok}, {Su}, {Marion}, \& {Marengo}}]{Banerjee+15}
{Banerjee}, D. P.~K., {Nuth}, Joseph~A., I., {Misselt}, K.~A., {et~al.} 2015,
  \apj, 814, 109, \dodoi{10.1088/0004-637X/814/2/109}

\bibitem[{{Baraffe} {et~al.}(2010){Baraffe}, {Chabrier}, \&
  {Barman}}]{Baraffe+10}
{Baraffe}, I., {Chabrier}, G., \& {Barman}, T. 2010, Reports on Progress in
  Physics, 73, 016901, \dodoi{10.1088/0034-4885/73/1/016901}

\bibitem[{{Belczynski} {et~al.}(2018){Belczynski}, {Askar}, {Arca-Sedda},
  {Chruslinska}, {Donnari}, {Giersz}, {Benacquista}, {Spurzem}, {Jin},
  {Wiktorowicz}, \& {Belloni}}]{Belczynski+18}
{Belczynski}, K., {Askar}, A., {Arca-Sedda}, M., {et~al.} 2018, \aap, 615, A91,
  \dodoi{10.1051/0004-6361/201732428}

\bibitem[{{Blagorodnova} {et~al.}(2017){Blagorodnova}, {Kotak}, {Polshaw},
  {et~al.}}]{Blagorodnova+17}
{Blagorodnova}, N., {Kotak}, R., {Polshaw}, J., {et~al.} 2017, \apj, 834, 107,
  \dodoi{10.3847/1538-4357/834/2/107}

\bibitem[{{Blagorodnova} {et~al.}(2021){Blagorodnova}, {Klencki}, {Pejcha},
  {Vreeswijk}, {Bond}, {Burdge}, {De}, {Fremling}, {Gehrz}, {Jencson},
  {Kasliwal}, {Kupfer}, {Lau}, {Masci}, \& {Rich}}]{Blagorodnova+21}
{Blagorodnova}, N., {Klencki}, J., {Pejcha}, O., {et~al.} 2021, \aap, 653,
  A134, \dodoi{10.1051/0004-6361/202140525}

\bibitem[{{Bond}(2019)}]{Bond19}
{Bond}, H.~E. 2019, \apj, 887, 12, \dodoi{10.3847/1538-4357/ab4e13}

\bibitem[{{Bond} {et~al.}(2003){Bond}, {Henden}, {Levay}, {Panagia}, {Sparks},
  {Starrfield}, {Wagner}, {Corradi}, \& {Munari}}]{Bond+03}
{Bond}, H.~E., {Henden}, A., {Levay}, Z.~G., {et~al.} 2003, \nat, 422, 405,
  \dodoi{10.1038/nature01508}

\bibitem[{{Cai} {et~al.}(2019){Cai}, {Pastorello}, {Fraser}, {et~al.}}]{Cai+19}
{Cai}, Y.~Z., {Pastorello}, A., {Fraser}, M., {et~al.} 2019, \aap, 632, L6,
  \dodoi{10.1051/0004-6361/201936749}

\bibitem[{{Chamandy} {et~al.}(2019){Chamandy}, {Tu}, {Blackman},
  {Carroll-Nellenback}, {Frank}, {Liu}, \& {Nordhaus}}]{Chamandy+19}
{Chamandy}, L., {Tu}, Y., {Blackman}, E.~G., {et~al.} 2019, \mnras, 486, 1070,
  \dodoi{10.1093/mnras/stz887}

\bibitem[{{Chevalier}(2012)}]{Chevalier12}
{Chevalier}, R.~A. 2012, \apjl, 752, L2, \dodoi{10.1088/2041-8205/752/1/L2}

\bibitem[{{Choi} {et~al.}(2016){Choi}, {Dotter}, {Conroy}, {Cantiello},
  {Paxton}, \& {Johnson}}]{Choi+16}
{Choi}, J., {Dotter}, A., {Conroy}, C., {et~al.} 2016, \apj, 823, 102,
  \dodoi{10.3847/0004-637X/823/2/102}

\bibitem[{{Chomiuk} {et~al.}(2021){Chomiuk}, {Metzger}, \& {Shen}}]{Chomiuk+21}
{Chomiuk}, L., {Metzger}, B.~D., \& {Shen}, K.~J. 2021, \araa, 59,
  \dodoi{10.1146/annurev-astro-112420-114502}

\bibitem[{{Clayton} {et~al.}(2017){Clayton}, {Podsiadlowski}, {Ivanova}, \&
  {Justham}}]{Clayton+17}
{Clayton}, M., {Podsiadlowski}, P., {Ivanova}, N., \& {Justham}, S. 2017,
  \mnras, 470, 1788, \dodoi{10.1093/mnras/stx1290}

\bibitem[{{Cummings} {et~al.}(2018){Cummings}, {Kalirai}, {Tremblay},
  {Ramirez-Ruiz}, \& {Choi}}]{Cummings+18}
{Cummings}, J.~D., {Kalirai}, J.~S., {Tremblay}, P.~E., {Ramirez-Ruiz}, E., \&
  {Choi}, J. 2018, \apj, 866, 21, \dodoi{10.3847/1538-4357/aadfd6}

\bibitem[{{Davies} {et~al.}(2004){Davies}, {Piotto}, \& {de
  Angeli}}]{Davies+04}
{Davies}, M.~B., {Piotto}, G., \& {de Angeli}, F. 2004, \mnras, 349, 129,
  \dodoi{10.1111/j.1365-2966.2004.07474.x}

\bibitem[{{De Marco} \& {Izzard}(2017)}]{DeMarco&Izzard17}
{De Marco}, O., \& {Izzard}, R.~G. 2017, \pasa, 34, e001,
  \dodoi{10.1017/pasa.2016.52}

\bibitem[{{de Mink} {et~al.}(2014){de Mink}, {Sana}, {Langer}, {Izzard}, \&
  {Schneider}}]{deMink+14}
{de Mink}, S.~E., {Sana}, H., {Langer}, N., {Izzard}, R.~G., \& {Schneider},
  F.~R.~N. 2014, \apj, 782, 7, \dodoi{10.1088/0004-637X/782/1/7}

\bibitem[{{Dexter} \& {Kasen}(2013)}]{Dexter&Kasen13}
{Dexter}, J., \& {Kasen}, D. 2013, \apj, 772, 30,
  \dodoi{10.1088/0004-637X/772/1/30}

\bibitem[{{Dotter}(2016)}]{Dotter16}
{Dotter}, A. 2016, \apjs, 222, 8, \dodoi{10.3847/0067-0049/222/1/8}

\bibitem[{{Faran} {et~al.}(2019){Faran}, {Goldfriend}, {Nakar}, \&
  {Sari}}]{Faran+19}
{Faran}, T., {Goldfriend}, T., {Nakar}, E., \& {Sari}, R. 2019, \apj, 879, 20,
  \dodoi{10.3847/1538-4357/ab218a}

\bibitem[{{Ferreira} {et~al.}(2019){Ferreira}, {Saito}, {Minniti}, {Navarro},
  {Ramos}, {Smith}, \& {Lucas}}]{Ferreira+19}
{Ferreira}, T., {Saito}, R.~K., {Minniti}, D., {et~al.} 2019, \mnras, 486,
  1220, \dodoi{10.1093/mnras/stz878}

\bibitem[{{Fragione} {et~al.}(2020){Fragione}, {Metzger}, {Perna}, {Leigh}, \&
  {Kocsis}}]{Fragione+20}
{Fragione}, G., {Metzger}, B.~D., {Perna}, R., {Leigh}, N. W.~C., \& {Kocsis},
  B. 2020, \mnras, 495, 1061, \dodoi{10.1093/mnras/staa1192}

\bibitem[{{Ge} {et~al.}(2020){Ge}, {Webbink}, \& {Han}}]{Ge+20}
{Ge}, H., {Webbink}, R.~F., \& {Han}, Z. 2020, \apjs, 249, 9,
  \dodoi{10.3847/1538-4365/ab98f6}

\bibitem[{{Glebbeek} {et~al.}(2013){Glebbeek}, {Gaburov}, {Portegies Zwart}, \&
  {Pols}}]{Glebbeek+13}
{Glebbeek}, E., {Gaburov}, E., {Portegies Zwart}, S., \& {Pols}, O.~R. 2013,
  \mnras, 434, 3497, \dodoi{10.1093/mnras/stt1268}

\bibitem[{{Gottlieb} {et~al.}(2022){Gottlieb}, {Tchekhovskoy}, \&
  {Margutti}}]{Gottlieb+22}
{Gottlieb}, O., {Tchekhovskoy}, A., \& {Margutti}, R. 2022, arXiv e-prints,
  arXiv:2201.04636.
\newblock \doarXiv{2201.04636}

\bibitem[{{Howitt} {et~al.}(2020){Howitt}, {Stevenson}, {Vigna-G{\'o}mez},
  {Justham}, {Ivanova}, {Woods}, {Neijssel}, \& {Mandel}}]{Howitt+20}
{Howitt}, G., {Stevenson}, S., {Vigna-G{\'o}mez}, A., {et~al.} 2020, \mnras,
  492, 3229, \dodoi{10.1093/mnras/stz3542}

\bibitem[{{Hurley} {et~al.}(2000){Hurley}, {Pols}, \& {Tout}}]{Hurley+00}
{Hurley}, J.~R., {Pols}, O.~R., \& {Tout}, C.~A. 2000, \mnras, 315, 543,
  \dodoi{10.1046/j.1365-8711.2000.03426.x}

\bibitem[{{Iben} \& {Livio}(1993)}]{Iben&Livio93}
{Iben}, Icko, J., \& {Livio}, M. 1993, \pasp, 105, 1373, \dodoi{10.1086/133321}

\bibitem[{{Ivanova} {et~al.}(2013{\natexlab{a}}){Ivanova}, {Justham}, {Avendano
  Nandez}, \& {Lombardi}}]{Ivanova+13b}
{Ivanova}, N., {Justham}, S., {Avendano Nandez}, J.~L., \& {Lombardi}, J.~C.
  2013{\natexlab{a}}, Science, 339, 433, \dodoi{10.1126/science.1225540}

\bibitem[{{Ivanova} \& {Nandez}(2016)}]{Ivanova&Nandez16}
{Ivanova}, N., \& {Nandez}, J.~L.~A. 2016, \mnras, 462, 362,
  \dodoi{10.1093/mnras/stw1676}

\bibitem[{{Ivanova} {et~al.}(2013{\natexlab{b}}){Ivanova}, {Justham}, {Chen},
  {De Marco}, {Fryer}, {Gaburov}, {Ge}, {Glebbeek}, {Han}, {Li}, {Lu}, {Marsh},
  {Podsiadlowski}, {Potter}, {Soker}, {Taam}, {Tauris}, {van den Heuvel}, \&
  {Webbink}}]{Ivanova+13a}
{Ivanova}, N., {Justham}, S., {Chen}, X., {et~al.} 2013{\natexlab{b}}, \aapr,
  21, 59, \dodoi{10.1007/s00159-013-0059-2}

\bibitem[{{Kankare} {et~al.}(2015){Kankare}, {Kotak}, {Pastorello},
  {et~al.}}]{Kankare+15}
{Kankare}, E., {Kotak}, R., {Pastorello}, A., {et~al.} 2015, \aap, 581, L4,
  \dodoi{10.1051/0004-6361/201526631}

\bibitem[{{Kasen} \& {Ramirez-Ruiz}(2010)}]{Kasen&RamirezRuiz10}
{Kasen}, D., \& {Ramirez-Ruiz}, E. 2010, \apj, 714, 155,
  \dodoi{10.1088/0004-637X/714/1/155}

\bibitem[{{Kasliwal} {et~al.}(2017){Kasliwal}, {Bally}, {Masci}, {Cody},
  {Bond}, {Jencson}, {Tinyanont}, {Cao}, {Contreras}, {et~al.}}]{Kasliwal+17}
{Kasliwal}, M.~M., {Bally}, J., {Masci}, F., {et~al.} 2017, \apj, 839, 88,
  \dodoi{10.3847/1538-4357/aa6978}

\bibitem[{{Kawash} {et~al.}(2021){Kawash}, {Chomiuk}, {Rodriguez}, {Strader},
  {Sokolovsky}, {Aydi}, {Kochanek}, {Stanek}, {Mukai}, {De}, {Shappee},
  {Holoien}, {Prieto}, \& {Thompson}}]{Kawash+21}
{Kawash}, A., {Chomiuk}, L., {Rodriguez}, J.~A., {et~al.} 2021, \apj, 922, 25,
  \dodoi{10.3847/1538-4357/ac1f1a}

\bibitem[{{Klencki} {et~al.}(2021){Klencki}, {Nelemans}, {Istrate}, \&
  {Chruslinska}}]{Klencki+21}
{Klencki}, J., {Nelemans}, G., {Istrate}, A.~G., \& {Chruslinska}, M. 2021,
  \aap, 645, A54, \dodoi{10.1051/0004-6361/202038707}

\bibitem[{{Kochanek} {et~al.}(2014){Kochanek}, {Adams}, \&
  {Belczynski}}]{Kochanek14}
{Kochanek}, C.~S., {Adams}, S.~M., \& {Belczynski}, K. 2014, \mnras, 443, 1319,
  \dodoi{10.1093/mnras/stu1226}

\bibitem[{{Kramer} {et~al.}(2020){Kramer}, {Schneider}, {Ohlmann}, {Geier},
  {Schaffenroth}, {Pakmor}, \& {R{\"o}pke}}]{Kramer+20}
{Kramer}, M., {Schneider}, F.~R.~N., {Ohlmann}, S.~T., {et~al.} 2020, \aap,
  642, A97, \dodoi{10.1051/0004-6361/202038702}

\bibitem[{{Kremer} {et~al.}(2022){Kremer}, {Lombardi}, {Lu}, {Piro}, \&
  {Rasio}}]{Kremer+22}
{Kremer}, K., {Lombardi}, James~C., J., {Lu}, W., {Piro}, A.~L., \& {Rasio},
  F.~A. 2022, arXiv e-prints, arXiv:2201.12368.
\newblock \doarXiv{2201.12368}

\bibitem[{{Kremer} {et~al.}(2021){Kremer}, {Lu}, {Piro}, {Chatterjee}, {Rasio},
  \& {Ye}}]{Kremer+21}
{Kremer}, K., {Lu}, W., {Piro}, A.~L., {et~al.} 2021, \apj, 911, 104,
  \dodoi{10.3847/1538-4357/abeb14}

\bibitem[{{Krishna Swamy}(1961)}]{KrishnaSwamy61}
{Krishna Swamy}, K.~S. 1961, \apj, 134, 1017, \dodoi{10.1086/147235}

\bibitem[{{Kruckow} {et~al.}(2016){Kruckow}, {Tauris}, {Langer}, {Sz{\'e}csi},
  {Marchant}, \& {Podsiadlowski}}]{Kruckow+16}
{Kruckow}, M.~U., {Tauris}, T.~M., {Langer}, N., {et~al.} 2016, \aap, 596, A58,
  \dodoi{10.1051/0004-6361/201629420}

\bibitem[{{Kulkarni} {et~al.}(2007){Kulkarni}, {Ofek}, {Rau}, {Cenko},
  {Soderberg}, {Fox}, {Gal-Yam}, {Capak}, {Moon}, {Li}, {Filippenko}, {Egami},
  {Kartaltepe}, \& {Sanders}}]{Kulkarni+07}
{Kulkarni}, S.~R., {Ofek}, E.~O., {Rau}, A., {et~al.} 2007, \nat, 447, 458,
  \dodoi{10.1038/nature05822}

\bibitem[{{Kurtenkov} {et~al.}(2015){Kurtenkov}, {Pessev}, {Tomov},
  {Barsukova}, {et~al.}}]{Kurtenkov+15}
{Kurtenkov}, A.~A., {Pessev}, P., {Tomov}, T., {Barsukova}, E.~A., {et~al.}
  2015, \aap, 578, L10, \dodoi{10.1051/0004-6361/201526564}

\bibitem[{{Law-Smith} {et~al.}(2020){Law-Smith}, {Everson}, {Ramirez-Ruiz}, {de
  Mink}, {van Son}, {G{\"o}tberg}, {Zellmann}, {Vigna-G{\'o}mez}, {Renzo},
  {Wu}, {Schr{\o}der}, {Foley}, \& {Hutchinson-Smith}}]{Law-Smith+20}
{Law-Smith}, J. A.~P., {Everson}, R.~W., {Ramirez-Ruiz}, E., {et~al.} 2020,
  arXiv e-prints, arXiv:2011.06630.
\newblock \doarXiv{2011.06630}

\bibitem[{{Lipunov} {et~al.}(2017){Lipunov}, {Blinnikov},
  {et~al.}}]{Lipunov+17}
{Lipunov}, V.~M., {Blinnikov}, S., {et~al.} 2017, \mnras, 470, 2339,
  \dodoi{10.1093/mnras/stx1107}

\bibitem[{{Lombardi} {et~al.}(2002){Lombardi}, {Warren}, {Rasio}, {Sills}, \&
  {Warren}}]{Lombardi+02}
{Lombardi}, James~C., J., {Warren}, J.~S., {Rasio}, F.~A., {Sills}, A., \&
  {Warren}, A.~R. 2002, \apj, 568, 939, \dodoi{10.1086/339060}

\bibitem[{{MacLeod} {et~al.}(2018{\natexlab{a}}){MacLeod}, {Cantiello}, \&
  {Soares-Furtado}}]{MacLeod+18b}
{MacLeod}, M., {Cantiello}, M., \& {Soares-Furtado}, M. 2018{\natexlab{a}},
  \apjl, 853, L1, \dodoi{10.3847/2041-8213/aaa5fa}

\bibitem[{{MacLeod} \& {Loeb}(2020)}]{MacLeod&Loeb20}
{MacLeod}, M., \& {Loeb}, A. 2020, \apj, 895, 29,
  \dodoi{10.3847/1538-4357/ab89b6}

\bibitem[{{MacLeod} {et~al.}(2017){MacLeod}, {Macias}, {Ramirez-Ruiz},
  {Grindlay}, {Batta}, \& {Montes}}]{MacLeod+17}
{MacLeod}, M., {Macias}, P., {Ramirez-Ruiz}, E., {et~al.} 2017, \apj, 835, 282,
  \dodoi{10.3847/1538-4357/835/2/282}

\bibitem[{{MacLeod} {et~al.}(2018{\natexlab{b}}){MacLeod}, {Ostriker}, \&
  {Stone}}]{MacLeod+18}
{MacLeod}, M., {Ostriker}, E.~C., \& {Stone}, J.~M. 2018{\natexlab{b}}, \apj,
  868, 136, \dodoi{10.3847/1538-4357/aae9eb}

\bibitem[{{Margalit} \& {Metzger}(2016)}]{Margalit&Metzger16}
{Margalit}, B., \& {Metzger}, B.~D. 2016, \mnras, 461, 1154,
  \dodoi{10.1093/mnras/stw1410}

\bibitem[{{Martini} {et~al.}(1999){Martini}, {Wagner}, {Tomaney}, {Rich},
  {della Valle}, \& {Hauschildt}}]{Martini+99}
{Martini}, P., {Wagner}, R.~M., {Tomaney}, A., {et~al.} 1999, \aj, 118, 1034,
  \dodoi{10.1086/300951}

\bibitem[{{Mason} {et~al.}(2010){Mason}, {Diaz}, {Williams}, {Preston}, \&
  {Bensby}}]{Mason+10}
{Mason}, E., {Diaz}, M., {Williams}, R.~E., {Preston}, G., \& {Bensby}, T.
  2010, \aap, 516, A108, \dodoi{10.1051/0004-6361/200913610}

\bibitem[{{Matsumoto} {et~al.}(2016){Matsumoto}, {Nakauchi}, {Ioka}, \&
  {Nakamura}}]{Matsumoto+16}
{Matsumoto}, T., {Nakauchi}, D., {Ioka}, K., \& {Nakamura}, T. 2016, \apj, 823,
  83, \dodoi{10.3847/0004-637X/823/2/83}

\bibitem[{{Mauerhan} {et~al.}(2015){Mauerhan}, {Van Dyk}, {Graham}, {Zheng},
  {Clubb}, {Filippenko}, {Valenti}, {Brown}, {Smith}, {Howell}, \&
  {Arcavi}}]{Mauerhan+15}
{Mauerhan}, J.~C., {Van Dyk}, S.~D., {Graham}, M.~L., {et~al.} 2015, \mnras,
  447, 1922, \dodoi{10.1093/mnras/stu2541}

\bibitem[{{Metzger} {et~al.}(2012){Metzger}, {Giannios}, \&
  {Spiegel}}]{Metzger+12}
{Metzger}, B.~D., {Giannios}, D., \& {Spiegel}, D.~S. 2012, \mnras, 425, 2778,
  \dodoi{10.1111/j.1365-2966.2012.21444.x}

\bibitem[{{Metzger} \& {Pejcha}(2017)}]{Metzger&Pejcha17}
{Metzger}, B.~D., \& {Pejcha}, O. 2017, \mnras, 471, 3200,
  \dodoi{10.1093/mnras/stx1768}

\bibitem[{{Metzger} {et~al.}(2017){Metzger}, {Shen}, \& {Stone}}]{Metzger+17}
{Metzger}, B.~D., {Shen}, K.~J., \& {Stone}, N. 2017, \mnras, 468, 4399,
  \dodoi{10.1093/mnras/stx823}

\bibitem[{{Metzger} {et~al.}(2021){Metzger}, {Zenati}, {Chomiuk}, {Shen}, \&
  {Strader}}]{Metzger+21}
{Metzger}, B.~D., {Zenati}, Y., {Chomiuk}, L., {Shen}, K.~J., \& {Strader}, J.
  2021, arXiv e-prints, arXiv:2108.04305.
\newblock \doarXiv{2108.04305}

\bibitem[{{Michaely} \& {Shara}(2021)}]{Michaely&Shara21}
{Michaely}, E., \& {Shara}, M.~M. 2021, \mnras, 502, 4540,
  \dodoi{10.1093/mnras/stab339}

\bibitem[{{Moe} \& {Di Stefano}(2017)}]{Moe&DiStefano17}
{Moe}, M., \& {Di Stefano}, R. 2017, \apjs, 230, 15,
  \dodoi{10.3847/1538-4365/aa6fb6}

\bibitem[{{Munari} {et~al.}(2002){Munari}, {Henden}, {Kiyota}, {Laney},
  {Marang}, {Zwitter}, {Corradi}, {Desidera}, {Marrese}, {Giro}, {Boschi}, \&
  {Schwartz}}]{Munari+02}
{Munari}, U., {Henden}, A., {Kiyota}, S., {et~al.} 2002, \aap, 389, L51,
  \dodoi{10.1051/0004-6361:20020715}

\bibitem[{{Nicholls} {et~al.}(2013){Nicholls}, {Melis}, {Soszynski}, {Udalski},
  {Szymanski}, {Kubiak}, {Pietrzynski}, {Poleski}, {Ulaczyk}, {Wyrzykowski},
  {Kozlowski}, \& {Pietrukowicz}}]{Nicholls+13}
{Nicholls}, C.~P., {Melis}, C., {Soszynski}, I., {et~al.} 2013, \mnras, 431,
  L33, \dodoi{10.1093/mnrasl/slt003}

\bibitem[{{Nordhaus} {et~al.}(2010){Nordhaus}, {Spiegel}, {Ibgui}, {Goodman},
  \& {Burrows}}]{Nordhaus+10}
{Nordhaus}, J., {Spiegel}, D.~S., {Ibgui}, L., {Goodman}, J., \& {Burrows}, A.
  2010, \mnras, 408, 631, \dodoi{10.1111/j.1365-2966.2010.17155.x}

\bibitem[{{Nordhaus} {et~al.}(2011){Nordhaus}, {Wellons}, {Spiegel}, {Metzger},
  \& {Blackman}}]{Nordhaus+11}
{Nordhaus}, J., {Wellons}, S., {Spiegel}, D.~S., {Metzger}, B.~D., \&
  {Blackman}, E.~G. 2011, Proceedings of the National Academy of Science, 108,
  3135, \dodoi{10.1073/pnas.1015005108}

\bibitem[{{Ohlmann} {et~al.}(2016){Ohlmann}, {R{\"o}pke}, {Pakmor}, \&
  {Springel}}]{Ohlmann+16}
{Ohlmann}, S.~T., {R{\"o}pke}, F.~K., {Pakmor}, R., \& {Springel}, V. 2016,
  \apjl, 816, L9, \dodoi{10.3847/2041-8205/816/1/L9}

\bibitem[{{Paczy{\'n}ski}(1970)}]{Paczynski70}
{Paczy{\'n}ski}, B. 1970, \actaa, 20, 47

\bibitem[{{Paczynski}(1976)}]{Paczynski76}
{Paczynski}, B. 1976, in Structure and Evolution of Close Binary Systems, ed.
  P.~{Eggleton}, S.~{Mitton}, \& J.~{Whelan}, Vol.~73, 75

\bibitem[{{Padmanabhan}(2000)}]{Padmanabhan00}
{Padmanabhan}, P. 2000, {Theoretical astrophysics. Vol.1: Astrophysical
  processes}

\bibitem[{{Pastorello} {et~al.}(2019{\natexlab{a}}){Pastorello}, {Mason},
  {Taubenberger}, {Fraser}, {Cortini}, {Tomasella}, {Botticella}, {Elias-Rosa},
  {Kotak}, {Smartt}, {Benetti}, {et~al.}}]{Pastorello+19a}
{Pastorello}, A., {Mason}, E., {Taubenberger}, S., {et~al.} 2019{\natexlab{a}},
  \aap, 630, A75, \dodoi{10.1051/0004-6361/201935999}

\bibitem[{{Pastorello} {et~al.}(2019{\natexlab{b}}){Pastorello}, {Chen}, {Cai},
  {Morales-Garoffolo}, {Cano}, {Mason}, {Barsukova}, {Benetti}, {Berton},
  {Bose}, {Bufano}, {et~al.}}]{Pastorello+19b}
{Pastorello}, A., {Chen}, T.~W., {Cai}, Y.~Z., {et~al.} 2019{\natexlab{b}},
  \aap, 625, L8, \dodoi{10.1051/0004-6361/201935511}

\bibitem[{{Pastorello} {et~al.}(2021{\natexlab{a}}){Pastorello}, {Fraser},
  {Valerin}, {Reguitti}, {Itagaki}, {Ochner}, {Williams}, {Jones}, {Munday},
  {Smartt}, {Smith}, {Srivastav}, {Elias-Rosa}, {Kankare}, {Karamehmetoglu},
  {Lundqvist}, {Mazzali}, {Munari}, {Stritzinger}, {Tomasella}, {Anderson},
  {Chambers}, \& {Rest}}]{Pastorello+21}
{Pastorello}, A., {Fraser}, M., {Valerin}, G., {et~al.} 2021{\natexlab{a}},
  \aap, 646, A119, \dodoi{10.1051/0004-6361/202039952}

\bibitem[{{Pastorello} {et~al.}(2021{\natexlab{b}}){Pastorello}, {Valerin},
  {Fraser}, {Elias-Rosa}, {Valenti}, {Reguitti}, {Mazzali}, {Amaro}, {Andrews},
  {Dong}, {Jencson}, {Lundquist}, {et~al.}}]{Pastorello+21b}
{Pastorello}, A., {Valerin}, G., {Fraser}, M., {et~al.} 2021{\natexlab{b}},
  \aap, 647, A93, \dodoi{10.1051/0004-6361/202039953}

\bibitem[{{Patterson}(1984)}]{Patterson84}
{Patterson}, J. 1984, \apjs, 54, 443, \dodoi{10.1086/190940}

\bibitem[{{Pavlovskii} {et~al.}(2017){Pavlovskii}, {Ivanova}, {Belczynski}, \&
  {Van}}]{Pavlovskii+17}
{Pavlovskii}, K., {Ivanova}, N., {Belczynski}, K., \& {Van}, K.~X. 2017,
  \mnras, 465, 2092, \dodoi{10.1093/mnras/stw2786}

\bibitem[{{Paxton} {et~al.}(2011){Paxton}, {Bildsten}, {Dotter}, {Herwig},
  {Lesaffre}, \& {Timmes}}]{Paxton+11}
{Paxton}, B., {Bildsten}, L., {Dotter}, A., {et~al.} 2011, \apjs, 192, 3,
  \dodoi{10.1088/0067-0049/192/1/3}

\bibitem[{{Paxton} {et~al.}(2013){Paxton}, {Cantiello}, {Arras}, {Bildsten},
  {Brown}, {Dotter}, {Mankovich}, {Montgomery}, {Stello}, {Timmes}, \&
  {Townsend}}]{Paxton+13}
{Paxton}, B., {Cantiello}, M., {Arras}, P., {et~al.} 2013, \apjs, 208, 4,
  \dodoi{10.1088/0067-0049/208/1/4}

\bibitem[{{Paxton} {et~al.}(2015){Paxton}, {Marchant}, {Schwab}, {Bauer},
  {Bildsten}, {Cantiello}, {Dessart}, {Farmer}, {Hu}, {Langer}, {Townsend},
  {Townsley}, \& {Timmes}}]{Paxton+15}
{Paxton}, B., {Marchant}, P., {Schwab}, J., {et~al.} 2015, \apjs, 220, 15,
  \dodoi{10.1088/0067-0049/220/1/15}

\bibitem[{{Paxton} {et~al.}(2019){Paxton}, {Smolec}, {Schwab}, {Gautschy},
  {Bildsten}, {Cantiello}, {Dotter}, {Farmer}, {Goldberg}, {Jermyn}, {Kanbur},
  {Marchant}, {Thoul}, {Townsend}, {Wolf}, {Zhang}, \& {Timmes}}]{Paxton+19}
{Paxton}, B., {Smolec}, R., {Schwab}, J., {et~al.} 2019, \apjs, 243, 10,
  \dodoi{10.3847/1538-4365/ab2241}

\bibitem[{{Pejcha}(2014)}]{Pejcha14}
{Pejcha}, O. 2014, \apj, 788, 22, \dodoi{10.1088/0004-637X/788/1/22}

\bibitem[{{Pejcha} {et~al.}(2016{\natexlab{a}}){Pejcha}, {Metzger}, \&
  {Tomida}}]{Pejcha+16a}
{Pejcha}, O., {Metzger}, B.~D., \& {Tomida}, K. 2016{\natexlab{a}}, \mnras,
  455, 4351, \dodoi{10.1093/mnras/stv2592}

\bibitem[{{Pejcha} {et~al.}(2016{\natexlab{b}}){Pejcha}, {Metzger}, \&
  {Tomida}}]{Pejcha+16b}
---. 2016{\natexlab{b}}, \mnras, 461, 2527, \dodoi{10.1093/mnras/stw1481}

\bibitem[{{Pejcha} {et~al.}(2017){Pejcha}, {Metzger}, {Tyles}, \&
  {Tomida}}]{Pejcha+17}
{Pejcha}, O., {Metzger}, B.~D., {Tyles}, J.~G., \& {Tomida}, K. 2017, \apj,
  850, 59, \dodoi{10.3847/1538-4357/aa95b9}

\bibitem[{{Perets} {et~al.}(2016){Perets}, {Li}, {Lombardi}, \&
  {Milcarek}}]{Perets+16}
{Perets}, H.~B., {Li}, Z., {Lombardi}, James~C., J., \& {Milcarek}, Stephen~R.,
  J. 2016, \apj, 823, 113, \dodoi{10.3847/0004-637X/823/2/113}

\bibitem[{{Podsiadlowski} {et~al.}(2004){Podsiadlowski}, {Langer},
  {Poelarends}, {Rappaport}, {Heger}, \& {Pfahl}}]{Podsiadlowski+04}
{Podsiadlowski}, P., {Langer}, N., {Poelarends}, A.~J.~T., {et~al.} 2004, \apj,
  612, 1044, \dodoi{10.1086/421713}

\bibitem[{{Popov}(1993)}]{Popov93}
{Popov}, D.~V. 1993, \apj, 414, 712, \dodoi{10.1086/173117}

\bibitem[{{Prust} \& {Chang}(2019)}]{Prust&Chang19}
{Prust}, L.~J., \& {Chang}, P. 2019, \mnras, 486, 5809,
  \dodoi{10.1093/mnras/stz1219}

\bibitem[{{Reichardt} {et~al.}(2020){Reichardt}, {De Marco}, {Iaconi},
  {Chamandy}, \& {Price}}]{Reichardt+20}
{Reichardt}, T.~A., {De Marco}, O., {Iaconi}, R., {Chamandy}, L., \& {Price},
  D.~J. 2020, \mnras, 494, 5333, \dodoi{10.1093/mnras/staa937}

\bibitem[{{Reichardt} {et~al.}(2019){Reichardt}, {De Marco}, {Iaconi}, {Tout},
  \& {Price}}]{Reichardt+19}
{Reichardt}, T.~A., {De Marco}, O., {Iaconi}, R., {Tout}, C.~A., \& {Price},
  D.~J. 2019, \mnras, 484, 631, \dodoi{10.1093/mnras/sty3485}

\bibitem[{{Ricker} \& {Taam}(2012)}]{Ricker&Taam12}
{Ricker}, P.~M., \& {Taam}, R.~E. 2012, \apj, 746, 74,
  \dodoi{10.1088/0004-637X/746/1/74}

\bibitem[{{Sana} {et~al.}(2012){Sana}, {de Mink}, {de Koter}, {Langer},
  {Evans}, {Gieles}, {Gosset}, {Izzard}, {Le Bouquin}, \&
  {Schneider}}]{Sana+12}
{Sana}, H., {de Mink}, S.~E., {de Koter}, A., {et~al.} 2012, Science, 337, 444,
  \dodoi{10.1126/science.1223344}

\bibitem[{{Sand} {et~al.}(2020){Sand}, {Ohlmann}, {Schneider}, {Pakmor}, \&
  {R{\"o}pke}}]{Sand+20}
{Sand}, C., {Ohlmann}, S.~T., {Schneider}, F. R.~N., {Pakmor}, R., \&
  {R{\"o}pke}, F.~K. 2020, \aap, 644, A60, \dodoi{10.1051/0004-6361/202038992}

\bibitem[{{Schaerer} \& {Maeder}(1992)}]{Schaerer&Maeder92}
{Schaerer}, D., \& {Maeder}, A. 1992, \aap, 263, 129

\bibitem[{{Schneider} {et~al.}(2019){Schneider}, {Ohlmann}, {Podsiadlowski},
  {R{\"o}pke}, {Balbus}, {Pakmor}, \& {Springel}}]{Schneider+19}
{Schneider}, F. R.~N., {Ohlmann}, S.~T., {Podsiadlowski}, P., {et~al.} 2019,
  \nat, 574, 211, \dodoi{10.1038/s41586-019-1621-5}

\bibitem[{{Shara} \& {Shaviv}(1977)}]{Shara&Shaviv77}
{Shara}, M.~M., \& {Shaviv}, G. 1977, \mnras, 179, 705,
  \dodoi{10.1093/mnras/179.4.705}

\bibitem[{{Smith}(2014)}]{Smith14}
{Smith}, N. 2014, \araa, 52, 487, \dodoi{10.1146/annurev-astro-081913-040025}

\bibitem[{{Smith} {et~al.}(2016){Smith}, {Andrews}, {Van Dyk}, {Mauerhan},
  {Kasliwal}, {Bond}, {Filippenko}, {Clubb}, {Graham}, {Perley}, {Jencson},
  {Bally}, {Ubeda}, \& {Sabbi}}]{Smith+16}
{Smith}, N., {Andrews}, J.~E., {Van Dyk}, S.~D., {et~al.} 2016, \mnras, 458,
  950, \dodoi{10.1093/mnras/stw219}

\bibitem[{{Soberman} {et~al.}(1997){Soberman}, {Phinney}, \& {van den
  Heuvel}}]{Soberman+97}
{Soberman}, G.~E., {Phinney}, E.~S., \& {van den Heuvel}, E.~P.~J. 1997, \aap,
  327, 620.
\newblock \doarXiv{astro-ph/9703016}

\bibitem[{{Soker}(2020)}]{Soker20}
{Soker}, N. 2020, \apj, 893, 20, \dodoi{10.3847/1538-4357/ab7dbb}

\bibitem[{{Soker} {et~al.}(2018){Soker}, {Grichener}, \& {Sabach}}]{Soker+18}
{Soker}, N., {Grichener}, A., \& {Sabach}, E. 2018, \apjl, 863, L14,
  \dodoi{10.3847/2041-8213/aad736}

\bibitem[{{Soker} \& {Kaplan}(2021)}]{Soker&Kaplan21}
{Soker}, N., \& {Kaplan}, N. 2021, Research in Astronomy and Astrophysics, 21,
  090, \dodoi{10.1088/1674-4527/21/4/90}

\bibitem[{{Soker} \& {Kashi}(2016)}]{Soker&Kashi16}
{Soker}, N., \& {Kashi}, A. 2016, \mnras, 462, 217,
  \dodoi{10.1093/mnras/stw1686}

\bibitem[{{Soker} \& {Tylenda}(2006)}]{Soker&Tylenda06}
{Soker}, N., \& {Tylenda}, R. 2006, \mnras, 373, 733,
  \dodoi{10.1111/j.1365-2966.2006.11056.x}

\bibitem[{{Stritzinger} {et~al.}(2020){Stritzinger}, {Taddia}, {Fraser},
  {et~al.}}]{Stritzinger+20}
{Stritzinger}, M.~D., {Taddia}, F., {Fraser}, M., {et~al.} 2020, \aap, 639,
  A104, \dodoi{10.1051/0004-6361/202038019}

\bibitem[{{Strope} {et~al.}(2010){Strope}, {Schaefer}, \& {Henden}}]{Strope+10}
{Strope}, R.~J., {Schaefer}, B.~E., \& {Henden}, A.~A. 2010, \aj, 140, 34,
  \dodoi{10.1088/0004-6256/140/1/34}

\bibitem[{{Sukhbold} {et~al.}(2016){Sukhbold}, {Ertl}, {Woosley}, {Brown}, \&
  {Janka}}]{Sukhbold+16}
{Sukhbold}, T., {Ertl}, T., {Woosley}, S.~E., {Brown}, J.~M., \& {Janka}, H.~T.
  2016, \apj, 821, 38, \dodoi{10.3847/0004-637X/821/1/38}

\bibitem[{{Taam} {et~al.}(1978){Taam}, {Bodenheimer}, \& {Ostriker}}]{Taam+78}
{Taam}, R.~E., {Bodenheimer}, P., \& {Ostriker}, J.~P. 1978, \apj, 222, 269,
  \dodoi{10.1086/156142}

\bibitem[{{Tauris} {et~al.}(2017){Tauris}, {Kramer}, {Freire}, {Wex}, {Janka},
  {Langer}, {Podsiadlowski}, {Bozzo}, {Chaty}, {Kruckow}, {van den Heuvel},
  {Antoniadis}, {Breton}, \& {Champion}}]{Tauris+17}
{Tauris}, T.~M., {Kramer}, M., {Freire}, P.~C.~C., {et~al.} 2017, \apj, 846,
  170, \dodoi{10.3847/1538-4357/aa7e89}

\bibitem[{{Tout} {et~al.}(2008){Tout}, {Wickramasinghe}, {Liebert}, {Ferrario},
  \& {Pringle}}]{Tout+08}
{Tout}, C.~A., {Wickramasinghe}, D.~T., {Liebert}, J., {Ferrario}, L., \&
  {Pringle}, J.~E. 2008, \mnras, 387, 897,
  \dodoi{10.1111/j.1365-2966.2008.13291.x}

\bibitem[{{Tylenda}(2005)}]{Tylenda05}
{Tylenda}, R. 2005, \aap, 436, 1009, \dodoi{10.1051/0004-6361:20052800}

\bibitem[{{Tylenda} {et~al.}(2005){Tylenda}, {Soker}, \&
  {Szczerba}}]{Tylenda+05}
{Tylenda}, R., {Soker}, N., \& {Szczerba}, R. 2005, \aap, 441, 1099,
  \dodoi{10.1051/0004-6361:20042485}

\bibitem[{{Tylenda} {et~al.}(2011){Tylenda}, {Hajduk}, {Kami{\'n}ski},
  {Udalski}, {Soszy{\'n}ski}, {Szyma{\'n}ski}, {Kubiak}, {Pietrzy{\'n}ski},
  {Poleski}, {Wyrzykowski}, \& {Ulaczyk}}]{Tylenda+11}
{Tylenda}, R., {Hajduk}, M., {Kami{\'n}ski}, T., {et~al.} 2011, \aap, 528,
  A114, \dodoi{10.1051/0004-6361/201016221}

\bibitem[{{Vigna-G{\'o}mez} {et~al.}(2018){Vigna-G{\'o}mez}, {Neijssel},
  {Stevenson}, {Barrett}, {Belczynski}, {Justham}, {de Mink}, {M{\"u}ller},
  {Podsiadlowski}, {Renzo}, {Sz{\'e}csi}, \& {Mandel}}]{Vigna-Gomez+18}
{Vigna-G{\'o}mez}, A., {Neijssel}, C.~J., {Stevenson}, S., {et~al.} 2018,
  \mnras, 481, 4009, \dodoi{10.1093/mnras/sty2463}

\bibitem[{{Wang} {et~al.}(2020){Wang}, {Kroupa}, {Takahashi}, \&
  {Jerabkova}}]{Wang+20}
{Wang}, L., {Kroupa}, P., {Takahashi}, K., \& {Jerabkova}, T. 2020, \mnras,
  491, 440, \dodoi{10.1093/mnras/stz3033}

\bibitem[{{Williams} {et~al.}(2015){Williams}, {Darnley}, {Bode}, \&
  {Steele}}]{Williams+15}
{Williams}, S.~C., {Darnley}, M.~J., {Bode}, M.~F., \& {Steele}, I.~A. 2015,
  \apjl, 805, L18, \dodoi{10.1088/2041-8205/805/2/L18}

\bibitem[{{Wilson} \& {Nordhaus}(2019)}]{Wilson&Nordhaus19}
{Wilson}, E.~C., \& {Nordhaus}, J. 2019, \mnras, 485, 4492,
  \dodoi{10.1093/mnras/stz601}

\bibitem[{{Wisniewski} {et~al.}(2008){Wisniewski}, {Clampin}, {Bjorkman}, \&
  {Barry}}]{Wisniewski+08}
{Wisniewski}, J.~P., {Clampin}, M., {Bjorkman}, K.~S., \& {Barry}, R.~K. 2008,
  \apjl, 683, L171, \dodoi{10.1086/591735}

\bibitem[{{Woosley}(2019)}]{Woosley19}
{Woosley}, S.~E. 2019, \apj, 878, 49, \dodoi{10.3847/1538-4357/ab1b41}

\end{thebibliography}

\end{document}